\documentclass[twocolumn]{aastex63}
\hypersetup{linkcolor=red,citecolor=green,filecolor=cyan,urlcolor=magenta}
\usepackage{multirow}
\usepackage{amsmath}
\usepackage{amssymb}
\usepackage{balance}
\newcommand\goldnum{140 }

\accepted{July 6, 2021}

\shorttitle{Multiwavelength properties of Miras}
\shortauthors{Iwanek et al.}

\begin{document}

\title{Multiwavelength properties of Miras}

\correspondingauthor{Patryk Iwanek}
\email{piwanek@astrouw.edu.pl}

\author[0000-0002-6212-7221]{Patryk Iwanek}
\affiliation{Astronomical Observatory, University of Warsaw, Al. Ujazdowskie 4, 00-478 Warsaw, Poland}

\author[0000-0003-4084-880X]{Szymon Koz{\l}owski}
\affiliation{Astronomical Observatory, University of Warsaw, Al. Ujazdowskie 4, 00-478 Warsaw, Poland}

\author[0000-0002-1650-1518]{Mariusz Gromadzki}
\affiliation{Astronomical Observatory, University of Warsaw, Al. Ujazdowskie 4, 00-478 Warsaw, Poland}

\author[0000-0002-7777-0842]{Igor Soszy{\'n}ski}
\affiliation{Astronomical Observatory, University of Warsaw, Al. Ujazdowskie 4, 00-478 Warsaw, Poland}

\author[0000-0002-3051-274X]{Marcin Wrona}
\affiliation{Astronomical Observatory, University of Warsaw, Al. Ujazdowskie 4, 00-478 Warsaw, Poland}

\author[0000-0002-2335-1730]{Jan Skowron}
\affiliation{Astronomical Observatory, University of Warsaw, Al. Ujazdowskie 4, 00-478 Warsaw, Poland}

\author[0000-0002-3218-2684]{Milena Ratajczak}
\affiliation{Astronomical Observatory, University of Warsaw, Al. Ujazdowskie 4, 00-478 Warsaw, Poland}

\author[0000-0001-5207-5619]{Andrzej Udalski}
\affiliation{Astronomical Observatory, University of Warsaw, Al. Ujazdowskie 4, 00-478 Warsaw, Poland}

\author[0000-0002-0548-8995]{Michał K. Szymański}
\affiliation{Astronomical Observatory, University of Warsaw, Al. Ujazdowskie 4, 00-478 Warsaw, Poland}

\author[0000-0002-2339-5899]{Paweł Pietrukowicz}
\affiliation{Astronomical Observatory, University of Warsaw, Al. Ujazdowskie 4, 00-478 Warsaw, Poland}

\author[0000-0001-6364-408X]{Krzysztof Ulaczyk}
\affiliation{Department of Physics, University of Warwick, Coventry CV4 7 AL, UK}
\affiliation{Astronomical Observatory, University of Warsaw, Al. Ujazdowskie 4, 00-478 Warsaw, Poland}

\author[0000-0002-9245-6368]{Radosław Poleski}
\affiliation{Astronomical Observatory, University of Warsaw, Al. Ujazdowskie 4, 00-478 Warsaw, Poland}

\author[0000-0001-7016-1692]{Przemysław Mróz}
\affiliation{Division of Physics, Mathematics, and Astronomy, California Institute of Technology, Pasadena, CA 91125, USA}
\affiliation{Astronomical Observatory, University of Warsaw, Al. Ujazdowskie 4, 00-478 Warsaw, Poland}

\author[0000-0001-9439-604X]{Dorota M. Skowron}
\affiliation{Astronomical Observatory, University of Warsaw, Al. Ujazdowskie 4, 00-478 Warsaw, Poland}

\author[0000-0002-9326-9329]{Krzysztof Rybicki}
\affiliation{Astronomical Observatory, University of Warsaw, Al. Ujazdowskie 4, 00-478 Warsaw, Poland}

\begin{abstract}
We comprehensively study the variability of Miras in the Large Magellanic Cloud (LMC) by simultaneous analysing light curves in 14 bands in the range of 0.5--24 microns.
We model over 20-years-long, high cadence $I$-band light curves collected by The Optical Gravitational Lensing Experiment (OGLE) and fit them to light curves collected in the remaining optical/near-infrared/mid-infrared bands to derive both the variability amplitude ratio and phase-lag as a function of wavelength. We show that the variability amplitude ratio declines with the increasing wavelength for both oxygen-rich (O-rich) and carbon-rich (C-rich) Miras, while the variability phase-lag increases slightly with the increasing wavelength. In a significant number of Miras, mostly the C-rich ones, the spectral energy distributions (SEDs) require a presence of a cool component (dust) in order to match the mid-IR data. 
Based on SED fits for a golden sample of \goldnum Miras, we calculated synthetic period-luminosity relations (PLRs) in 42 bands for the existing and future sky surveys that include OGLE, The VISTA Near-Infrared $YJK_\mathrm{s}$ Survey of the Magellanic Clouds System (VMC), Legacy Survey of Space and Time (LSST), Gaia, Spitzer, The Wide-field Infrared Survey Explorer (WISE), The James Webb Space Telescope (JWST), The Nancy Grace Roman Space Telescope (formerly WFIRST), and The Hubble Space Telescope (HST). We show that the synthetic PLR slope decreases with increasing wavelength for both the O-rich and C-rich Miras in the range of 0.1--40 microns. Finally, we show the location and motions of Miras on the color-magnitude (CMD) and color-color (CCD) diagrams. 
\end{abstract}

\keywords{stars: AGB and post-AGB -- stars: carbon -- stars: variables: general -- Magellanic Clouds}


\section{Introduction} \label{sec:introduction}

The Mira star ($o$ Ceti) is historically the first identified variable star in modern astronomy. The discovery of its variability by David Fabricius at the end of the 16th century, 
and periodicity by Johannes Holwards in the 17th century has led to the birth of one of the most important branches of astronomical research -- stellar variability.

The Mira-type stars are fundamental-mode Asymptotic Giant Branch (AGB) pulsators belonging to a group of the Long Period Variables (LPVs). The pulsation periods of Miras span a range between $\sim$100 days and 1000 days, or more. The Mira-type variables are relatively easy to detect due to large brightness variations in optical bands with $\Delta V > 2.5$ mag (e.g., \citealt{1951astr.conf..495P, 2017ARep...61...80S}), $\Delta I > 0.8$ mag (e.g., \citealt{2005AcA....55..331S, 2009AcA....59..239S}), and the decreasing variability amplitude at longer wavelengths ($\Delta K > 0.4$ mag; e.g., \citealt{2006MNRAS.369..751W}). The upper boundary of brightness variations in $K$-band seems to be about $1$~mag \citep{1982MNRAS.201..439F}.

During their evolution through the tip of the AGB, Miras become very luminous (several thousand $L_\odot$). Due to their large luminosity, Mira-type stars are tracers of an old- and intermediate-age stellar populations in many galaxies, e.g., in the Large Magellanic Cloud \citep[LMC,][]{2009AcA....59..239S}, Small Magellanic Cloud \citep[SMC,][]{2011AcA....61..217S}, NGC 6822 \citep{2013MNRAS.428.2216W}, IC~1613 \citep{2015MNRAS.452..910M}, M33 \citep{2017AJ....153..170Y}, Sgr~dIG \citep{2018MNRAS.473..173W}, NGC~4258 \citep{2018ApJ...857...67H}, NGC 3109 \citep{2019MNRAS.483.5150M}, or NGC~1559 \citep{2020ApJ...889....5H}. Moreover, Miras obey well-defined PLRs in the near-infrared (NIR) and mid-infrared wavelengths (mid-IR, e.g., \citealt{2011MNRAS.412.2345I, 2018AJ....156..112Y, 2019ApJ...884...20B, 2020A&A...636A..48G, 2020arXiv201212910I}), so they can be used as distance indicators.

The AGB stars are typically divided into the oxygen-rich (O-rich) and carbon-rich (C-rich) classes (see, e.g., \citealt{2010ApJ...723.1195R}). Such a division has been systematized by \cite{2009AcA....59..239S} for the LMC LPVs. The authors divided LPVs into O-rich and C-rich stars based on the color-color ($(V-I)$ vs. $(J-K)$) diagram and Wesenheit diagram ($W_{I}$ vs. $W_{JK}$). Other authors proposed a similar division for the LPVs located in the Milky Way \citep{2020ApJS..247...44A} or nearby galaxies (e.g., \citealt{2017AJ....153..170Y,2018AJ....156..112Y,2018ApJ...857...67H}). 

Mira variables, as luminous high-amplitude AGB pulsators, are suspected to undergo the mass-loss phenomenon via the stellar wind  \citep{2020A&A...642A..82P}. Dying stars during the AGB phase eject a large amount of heavy elements created in a slow neutron-capture process \mbox{(s-process)}, which leads to an enrichment of the interstellar medium with elements heavier than nickel. This phenomenon has been extensively studied due to its importance from the point of view of chemistry of galaxies (e.g., \citealt{1997MNRAS.288..512W, 2016A&A...586A..49B, 2017A&A...598A..53D, 2018A&ARv..26....1H, 2019ApJ...887...41L, 2019ApJ...887...82K, 2021MNRAS.501.5135Y}).

The variability of Miras, like other pulsators, is not strictly periodic. The long-term changes of the mean brightness (possibly caused by dust ejections), and cycle-to-cycle light and period variations have been observed in the past decades (e.g., \citealt{1929MNRAS..90...65E, 1937AnHar.105..459S, 1989Obs...109..146L, 1997JAVSO..25...57M, 1999PASP..111...98P}). \citet{1984MNRAS.211..331F} discovered the variable dust obscuration around C-rich Mira -- R Fornacis. A decade later, \citet{1994A&A...290..623W} showed synthetic mid-IR light curves for C-rich stars, based on dynamical models of circumstellar dust shells. The authors showed models for R Fornacis along with data collected by \citet{1984MNRAS.211..331F}.
The variability amplitude ratio between the $KLM$- and $J$-band light curves decreases with increasing wavelength. The variability amplitude ratio presented by \citet{1994A&A...290..623W} are $\sim 0.61$ and $\sim 0.42$ for the $K$/$J$- and $L$/$J$-bands, respectively. From the theoretical light curves, it was shown that the variability amplitude ratio between the $25$ micron and $J$-band light curves is approximately equal to $0.1$. A phase-shift between the $KLM$- and $J$-band light curves for R Fornacis was not reported.

\citet{2003MNRAS.342...86W} noticed that pulsation amplitudes of Miras, as well as amplitudes of their long-term trends, are smaller at longer wavelengths. \citet{2005A&A...441.1117L} reported, that the pulsation amplitude ratio between the $K$- and $V$-band is equal to $0.2$. Recently, \citet{2018AJ....156..112Y} presented the variability amplitude ratio between the $JHK_\mathrm{s}$-bands and $I$-band, separately for the \mbox{O-rich} (0.415, 0.435, 0.409, respectively) and C-rich (0.771, 0.660, 0.538, respectively) Miras. These ratios differ significantly for the two sub-classes, and are larger for the C-rich Miras. 

Moreover, it is known that the variability maxima of Mira light curves are delayed in the near-IR with respect to the visual bands by about 0.1 period \citep{1933ApJ....78..320P, 2006AJ....131..612S}. These phase-lags were also studied by \citet{2018AJ....156..112Y}. In contrast to the variability amplitude ratio, the phase-lag between the $I$-band and $JHK_\mathrm{s}$-bands is larger for the O-rich Miras and increases with the increasing wavelength. The mean phase-lags between the $I$-band and $JHK_\mathrm{s}$-bands for the O-rich Miras are $0.126$, $0.133$, and $0.155$, respectively. The phase-lag observed in the C-rich Miras is the largest in the $K_\mathrm{s}$-band, and is equal to $0.03$ phase. Similar conclusions were made by \citet{2021MNRAS.500...82I}.

The aim of this paper is to comprehensively analyze the variability of Miras at a wide range of optical/near-IR/mid-IR wavelengths. 
This work contains two parts. The first one delivers analyses of variability based on the real data in up to 14 filters, while the second part provides analyses based on synthetic light curves derived from spectral energy distribution (SED) fitted to the real data in the 0.1--40 micron range. We derive the variability amplitude ratio and the phase-lag as a function of wavelength for both the real and synthetic data.

In Section~\ref{sec:data}, we describe the data used in this work that include the sample of Miras and extinction correction procedure, and
in Section~\ref{sec:methods} we describe the light curve modeling and fitting.
We present and discuss the variability amplitude ratio and the phase-lag as a function of wavelength in Section~\ref{sec:varpha}.
In Section~\ref{sec:SED}, we discuss the synthetic data that include the SED fitting and synthetic PLRs. The Section \ref{sec:CMDs} is devoted to location and motions of Miras on the color-magnitude (CMD) and color-color (CCD) diagrams. The paper is summarized in Section~\ref{sec:summary}, while in Section \ref{sec:future}, we outline future areas of exploration.

\newpage
\section{Data} 
\label{sec:data}

\subsection{Sample of Miras} \label{subsec:mirassample}

In this study, we used the sample of $1663$ Miras discovered in the LMC by the OGLE during third phase of the project \citep[OGLE-III][]{2003AcA....53..291U, 2005AcA....55..331S,2009AcA....59..239S}. The authors discovered almost $100\;000$ LPVs and provided two-band light curves (\mbox{$I$- and} $V$-bands from the Johnson-Cousins photometric system) spanning $\sim 13$ years since $1996$. LPVs in that catalog were divided into three subclasses: {\it OGLE Small Amplitude Red Giants} (OSARGs), {\it Semi Regular Variables} (SRVs) and Miras. The two latter subclasses were separated using the $I$-band pulsation amplitude with $\Delta I > 0.8$ mag for Miras. \citet{2009AcA....59..239S} divided AGB stars into the O-rich and C-rich classes using the color-color diagram ($(V-I)$ vs. $(J-K)$) as well as the Wesenheit diagram ($W_I$ vs. $W_{JK}$). As a result, $1194$ C-rich Miras, and $469$ O-rich Miras were classified. This division was evaluated and confirmed by \citet{2011MNRAS.412.2345I}. 
We use this classification throughout this paper.
The full catalog containing pulsation periods, coordinates, surfaces chemistry classification and many more information is publicly available and can be access through the OGLE webpage\footnote{http://ogle.astrouw.edu.pl}.

\subsection{Optical photometry} \label{subsec:ogle}
The OGLE project has started its operation in 1992. Since then, the Galactic bulge (BLG), and later also the Magellanic Clouds (MCs), and the Galactic disk (GD) have been regularly monitored in the $V$-band ($\lambda_{\mathrm{eff}} = 0.55$ $\mu$m) and $I$-band ($\lambda_{\mathrm{eff}} = 0.81$ $\mu$m) to search for stellar variability. The OGLE-III phase ended in 2009. In 2010, the Warsaw telescope, located at the Las Campanas Observatory, Chile, was equipped with a 32-chip CCD camera, what began the fourth phase of the Optical Gravitational Lensing Experiment project \citep[OGLE-IV,][]{2015AcA....65....1U}. The observations restarted in March, 2010. The monitoring of the southern sky lasted until March 17, 2020, when the OGLE-IV monitoring program had to be interrupted due to the COVID-19 pandemic.

To date the OGLE project increased the baseline of the $V$- and $I$-band Mira light curves from the LPVs catalog \citep{2009AcA....59..239S} by additional 11 years. We combined the \mbox{OGLE-II}, \mbox{OGLE-III}, and \mbox{OGLE-IV} light curves, making our sample of Miras the largest one with such a long time-span covering over two decades for the vast majority of objects.

In this paper, we revised the pulsation periods of Miras using the full, two-decades-long light curves. We used the {\sc TATRY} code, which implements the multiharmonic analysis of the variance algorithm \citep[ANOVA,][]{1996ApJ...460L.107S}. Each phase-folded light curve was inspected visually, and if necessary, the period was corrected manually. In some cases, manual corrections were required as Miras exhibit aperiodic amplitude and mean magnitude variations, that may lead to slight offsets in the detected periods. A visual inspection also helped to uncover and correct alias periods. The detailed discussion about selection of the period-searching method for unevenly sampled data can be found in \citet{2019ApJ...879..114I}. During the visual inspection of light curves, we removed significant and obvious outliers due to photometric problems. The earliest observation in the optical OGLE light curves was taken on 29th of December, 1996, and the last one was taken on 15th March, 2020. The median number of data points per light curve is $1310$ and $110$ in $I$-band and $V$-band, respectively. Both dates of the first/last observations and the number of epochs can vary between OGLE fields.

The catalog of the OGLE LPVs \citep{2009AcA....59..239S} will be updated in the near future by presently found periods, over two-decades-long light curves, but also by new Miras discoveries located outside the OGLE-III fields.

\subsection{NIR photometry} \label{subsec:vmc}

In our study, we also used measurements taken in the near-infrared $Y$-band ($\lambda_{\mathrm{eff}} = 1.02$ $\mu$m), $J$-band ($\lambda_{\mathrm{eff}} = 1.22$ $\mu$m), and $K_{S}$-band ($\lambda_{\mathrm{eff}} = 2.19$ $\mu$m) by the Near-Infrared $YJK_\mathrm{s}$ Survey of the Magellanic Clouds System \citep[VMC,][]{2011A&A...527A.116C}. These data were downloaded using Table Access Protocol (TAP) Query via TOPCAT\footnote{http://www.star.bris.ac.uk/$\sim$mbt/topcat/} \citep{2005ASPC..347...29T} from the ESO \texttt{tap\_cat} service. We cross-matched our sample with the epoch-merged and band-merged master source catalogs in $YJK_{s}$-bands (\texttt{vmc\_er4\_ksjy\_catMetaData\_fits\_V3}).
We found 595 (175 O-rich Miras, and 420 \mbox{C-rich} Miras) matches within 1 arscec radius.
Then, we retrieved the photometric data for these objects  using their \texttt{IAUNAME} from the multi-epoch $YJK_{s}$-bands catalogues: \texttt{vmc\_er4\_y\_mPhotMetaData\_fits\_V3}, 
\texttt{vmc\_er4\_j\_mPhotMetaData\_fits\_V3}
and
\newline \texttt{vmc\_er4\_ks\_mPhotMetaData\_fits\_V3}, respectively. The VMC observations were taken between November 2009 and August 2013 with at least $3$, $3$, and $12$ epochs in $YJK_{S}$-bands, respectively.   

\subsection{MID-IR photometry}  \label{subsec:midir}

\subsubsection{WISE data}  \label{subsubsec:wise}
We cross-matched the sample of LMC Miras with the {\it Wide Field Infrared Survey Explorer} \citep[WISE,][]{2010AJ....140.1868W} databases. WISE is a 40-cm diameter infrared space telescope that observed the sky in four bands: W1 ($\lambda_{\mathrm{eff}} = 3.4$ $\mu$m), W2 ($\lambda_{\mathrm{eff}} = 4.6$ $\mu$m), W3 ($\lambda_{\mathrm{eff}} =12$ $\mu$m) and W4 ($\lambda_{\mathrm{eff}} = 22$ $\mu$m). The main mission of the WISE telescope was to map the entire sky in these four mid-IR bands. This mission was completed in 2010. In early 2011, due to the depletion of the solid hydrogen cryostat, WISE was placed into hibernation mode. Re-observations began in 2013 as a part of the {\it Near Earth Object WISE Reactivation Mission} \citep[NEOWISE-R,][]{2011ApJ...731...53M, 2014ApJ...792...30M}, and have been carried out in $W1-$ and $W2-$bands to this day.

The WISE telescope observes each sky location every six months, and during one flyby, it collects several independent exposures, which span from one, to a few days. The LMC area (near the South Ecliptic Pole) are observed much more often due to the polar trajectory of the WISE telescope. As a result, light curves of the LMC stars are covered much more densely than in other locations.

We searched for objects around LMC Miras coordinates within $1$ arcsec. We downloaded all available measurements in the $W1$-, $W2$-, $W3$-, and $W4$-bands from the AllWISE Multiepoch Photometry Table\footnote{https://wise2.ipac.caltech.edu/docs/release/allwise/}, and all measurements in the $W1$- and $W2$-bands from the NEOWISE-R Single Exposure (L1b) Source Table using NASA/IPAC Infrared Science Archive\footnote{https://irsa.ipac.caltech.edu/applications/Gator/}. We found in the AllWISE table counterparts to $1592$, $1594$, $1616$, $1551$ (out of $1663$) Miras in the $W1$-, $W2$-, $W3$-, $W4$-bands, respectively. In the NEOWISE-R table, we found counterparts to $1632$, and $1643$ objects in the $W1$- and $W2$-bands, respectively.

\begin{figure}[ht!]
\centering
\includegraphics[scale=0.24]{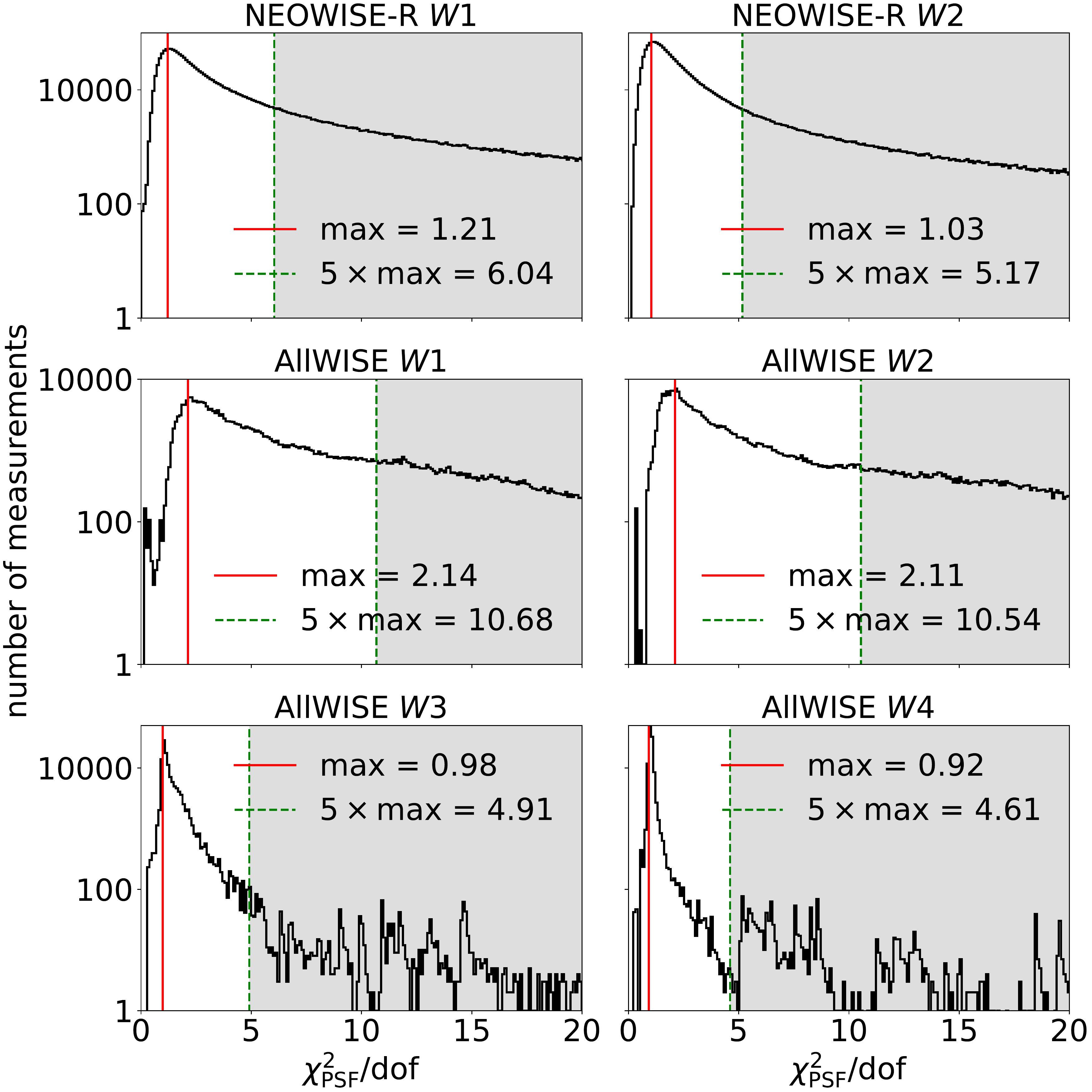}
\caption{Distributions of $\chi^2_{\mathrm{PSF}}/{\mathrm{dof}}$ from fitting the point spread function (PSF) to images of the WISE survey. As part of our two-step procedure of pruning the data from the outliers, we take into consideration points in the white area, located within five times of the $\chi^2_{\mathrm{PSF}}/{\mathrm{dof}}$ maximum. }
\label{fig:wise_chi2}
\end{figure}

For each single measurement in the database, the WISE team provided the reduced $\chi^2_{\mathrm{PSF}}$ from fitting the point spread function (PSF) to objects detected in the collected images. We examined the distributions of $\chi^2_{\mathrm{PSF}}/{\mathrm{dof}}$ in each WISE band, both in AllWISE and NEOWISE-R data.
We cleaned the WISE light curves from the worst-quality points, assuming that best-quality measurements have \mbox{$\chi^2_{\mathrm{PSF}}/{\mathrm{dof}} < 5 \times \chi^2_{\mathrm{PSF, max}}/{\mathrm{dof}}$}, where $\chi^2_{\mathrm{PSF, max}}/{\mathrm{dof}}$ is the highest frequencies of the $\chi^2_{\mathrm{PSF}}/{\mathrm{dof}}$ distributions in each band. The $\chi^2_{\mathrm{PSF}}/{\mathrm{dof}}$ distributions are presented in Figure \ref{fig:wise_chi2}. We rejected in total $24.6\%$, $15.6\%$, $2.3\%$, and $1.7\%$ observations in the $W1$, $W2$, $W3$, and $W4$ bands, respectively.

\vspace{0.2cm}

The WISE light curves are divided into ``epochs'', containing several dozen measurements each. The $W1$, $W2$, $W3$, and $W4$ AllWISE light curves comprise of one or two epochs only, while the $W1$ and $W2$ \mbox{NEOWISE-R} light curves contain at least 12 such epochs, separated by roughly half a year. We calculated the weighted mean magnitude in each epoch, and we rejected measurements deviating more than $3\sigma$ from the mean. After the visual inspection, we decided to use the NEOWISE-R measurements in the W1 and W2 bands only because of two reasons: the time-span was significantly longer and the number of data points was higher in the light curves retrieved from the NEOWISE-R table, and the internal scatter in the AllWISE and NEOWISE-R are different. We used the $W3$ and $W4$ light curves from the AllWISE database.

Finally, we were left with fully-cleaned light curves. The median number of data points per light curve were $645$, $703$, $63$, and $22$ in $W1$, $W2$, $W3$, and $W4$ bands, respectively. Light curves with the greatest coverage contained over $2500$ data points, and over $200$ data points for $W1$/$W2$, and $W3$/$W4$ bands, respectively. The AllWISE observations were taken between 8th February, 2010 and 17th June, 2010, while the NEOWISE-R data were collected between December 13th, 2010, and December 1st, 2019.

We removed from further analysis stars that had less than $100$ measurements in $W1$ and $W2$ bands, and less than $3$ data points in W3 and W4 bands. The final sample of Miras in the WISE bands contained 1311 stars (out of 1663).

\subsubsection{Spitzer data}  \label{subsubsec:spitzer}

The Spitzer Space Telescope is an 85-cm diameter telescope with three infrared instruments onboard: {\it Infrared Array Camera} (IRAC), {\it Infrared Spectrograph} (IRS), and {\it Multiband Imaging Photometry for Spitzer} (MIPS) \citep{2004ApJS..154....1W}. The LMC was observed by Spitzer in the {\it Surveying the Agents of Galaxy's Evolution} \citep[SAGE,][]{2006AJ....132.2268M} survey using both the IRAC and MIPS instruments. The IRAC is equipped with four channels: $[3.6]$ ($\lambda_{\mathrm{eff}} = 3.6$ $\mu$m), $[4.5]$ ($\lambda_{\mathrm{eff}} = 4.5$ $\mu$m), $[5.8]$ ($\lambda_{\mathrm{eff}} = 5.8$ $\mu$m), $[8.0]$ ($\lambda_{\mathrm{eff}} = 8.0$ $\mu$m), while MIPS is equipped with three channels: $[24.0]$ ($\lambda_{\mathrm{eff}} = 24$ $\mu$m), $[70.0]$ ($\lambda_{\mathrm{eff}} = 70$ $\mu$m), $[160.0]$ ($\lambda_{\mathrm{eff}} = 160$ $\mu$m). In this paper, we use $[3.6]$, $[4.5]$, $[5.8]$, $[8.0]$, $[24.0]$ channels, later also refereed as $I1$, $I2$, $I3$, $I4$, and $M1$, respectively.

We searched for objects within 1 arcsec radius around Miras coordinates. We downloaded all available measurements from the SAGE IRAC Epoch 1 and Epoch 2 Catalog using NASA/IPAC Infrared Science Archive\footnote{https://irsa.ipac.caltech.edu/applications/Gator/}. We found 1601 (out of 1663) counterparts to our Mira sample. The measurements provided in the SAGE IRAC Epoch 1 and Epoch 2 Catalog do not contain time of observations. Times for IRAC observations are available from the time stamp mosaic images\footnote{https://irsa.ipac.caltech.edu/data/SPITZER/SAGE/timestamps/}. The SAGE team provided {\it Interactive Data Language} (IDL) program \texttt{get\_sage\_jd.pro}. This program requires IDL procedures from the IDL Astronomy User's Library\footnote{https://idlastro.gsfc.nasa.gov; In case the user does not have access to the IDL interpreter, all routines and \texttt{get\_sage\_jd.pro} program could be use with free alternative to IDL --  GNU Data Language \citep[GDL,][]{2010ASPC..434..187C}.}. Unfortunately, the time stamps for the MIPS observations are not publicly available.

Almost all SAGE IRAC light curves contain 2 epochs, collected from October to November 2005. For further analysis we used objects, with 2 epochs in each IRAC band. This left us with 1470 Miras (out of 1663).

Additionally, we retrieved data from the SAGE MIPS 24 $\mu$m Epoch 1 and Epoch 2 Catalog. We found 1415 counterparts to our objects. As the observation time is not publicly available for MIPS measurements, we used this data in further analysis, treating them as a mean.

\section{Methods}
\label{sec:methods}

The main aim of this paper is to study the variability of Miras spanning a range of wavelengths, and then to find the variability amplitude ratios and phase-lags for these stars at a number of wavelengths. 

\subsection{Extinction corrections} 
\label{sec:extinction}

In this paper, we are interested in analyzing Miras located in the LMC, therefore their light is subject to absorption by interstellar dust. The most detailed reddening map of the LMC is based on LMC Red Clump (RC) stars, and was published by \citet[][hereafter S21]{2021ApJS..252...23S}. The authors provided the reddening $E(V-I)$ coefficients, along with the lower ($\sigma_{-,E(V-I)}$) and upper ($\sigma_{+,E(V-I)}$) uncertainties. For a given object, the reddening $E(V-I)$ could be retrieved from the on-line form\footnote{http://ogle.astrouw.edu.pl/cgi-ogle/get\_ms\_ext.py} by using their Right Ascension (RA) and Declination (Decl.). The extinction $A_I$ could be calculated as:

\begin{equation}
    A_I \simeq 1.5 \times E(V-I),
    \label{eqn:A_I}
\end{equation}

\noindent where the coefficient depends on the inner LMC dust characteristics, and could vary between $1.1$ and $1.7$. We fixed this coefficient to $1.5$ with the uncertainty equal to $0.2$. Knowing that $E(V-I) = A_V - A_I$, the extinction $A_V$ is:

\begin{equation}
    A_V \simeq 2.5 \times E(V-I).
    \label{eqn:A_V}
\end{equation}

\noindent This is broadly consistent with coefficients 1.505 for $A_I$ and 2.742 for $A_V$ derived for $R_V = A_V/E(B-V) = 3.1$ by \cite{2011ApJ...737..103S}.
Both extinctions $A_V$ and $A_I$ can be transformed to other bands using relations published by \citet{2019ApJ...877..116W}.

As a cross-check, we compared the extinctions obtained with the S21 method to the extinction law derived by \citet{1989ApJ...345..245C}. We first calculated the extinction in the K-band $A_K = 0.219 \times E(V-I)$, using the reddening $E(V-I)$ provided by \citet{2021ApJS..252...23S}, and then we transformed $A_K$ to other bands using relations published by \citet{Chen_2018}. This method will be referred to as C89.
We then compared the extinction obtained using both methods for two IR bands: $J$ ($1.25$ $\mu$m) and $I4$ ($8.0$ $\mu$m). 

The extinction toward the LMC at short wavelengths is relatively small (the median value in $J$-band is $A_J = 0.07$ mag). Then, the influence of interstellar dust on stellar light decreases with increasing 
wavelength (see e.g., \citealt{1989ApJ...345..245C}). In the Spitzer $I4$-band, the extinction is one order of magnitude smaller than in $J-$band (the median value is $A_{I4} = 0.007$ mag). In general, in the $I4$-band the extinction is comparable to the level of a single photometric measurement uncertainty for the most surveys. The difference between the S21 and C89 methods is less than $10\%$.

Throughout the paper, we used the S21 method to correct the magnitudes for the interstellar extinction, and both the stellar brightness and colors are extinction corrected and dereddened.

\subsection{Template light curves} \label{subsec:templates}

The variability of Miras is characterized by several components and is not strictly periodic, therefore their light curves can not be adequately modeled with a pure periodic function (e.g., a sine wave). The main large-amplitude, cyclic variability of Miras is caused by pulsations \citep{1985IBVS.2681....1K}. \citet{1997JAVSO..25...57M} noticed cycle-to-cycle and long-term magnitude changes, which are associated with the mass-loss and presence of circumstellar dust. The last, stochastic component of Miras variability is related to supergranular convection in the envelopes of giants stars.

The semi-parametric Gaussian process regression (GPR) model, which takes into account all these types of brightness changes, was proposed by \citet{2016AJ....152..164H}. The Miras light curve signal $g(t)$ can be decomposed into four parts:

\begin{equation}
    g(t)  = m + l(t) + q(t) + h(t),
\label{eqn:model}
\end{equation}

\noindent where $m$ is the mean magnitude, $l(t)$ is low-frequency trend across cycles (or slowly-variable mean magnitude), $q(t)$ is periodic term, and $h(t)$ is high-frequency, stochastic variability within each cycle. The $l(t)$, $q(t)$, and $h(t)$ terms are modeled by the Gaussian process with different kernels (in particular squared exponential kernels, and periodic kernel). The use of the periodic kernel allows the brightness amplitude to change from cycle to cycle, what is a characteristic feature of Miras. Such a complex variability behavior of Miras is not reproducible by strictly periodic functions. The full code is distributed as an \mbox{R \citep{Rsoftware}} package via GitHub \citep{GPR}. Using the above model and equipped with the high-quality, two-decade-long, and densely sampled OGLE $I$-band light curves, we derived template light curves for the selected LMC Miras. Both the dense sampling and high precision of OGLE $I$-band light curves lead to very smooth template light curves. Given that the GPR model is data-driven, the GPR models describe not only the $I$-band light curves exceptionally well, but also scaled and shifted template fits to other bands, in particular to the $V$-band light curves (in Figures~\ref{fig:lc_o} and \ref{fig:lc_c}). The synthetic light curve can be generated with any cadence.

\begin{figure*}
\centering
\includegraphics[scale=0.6]{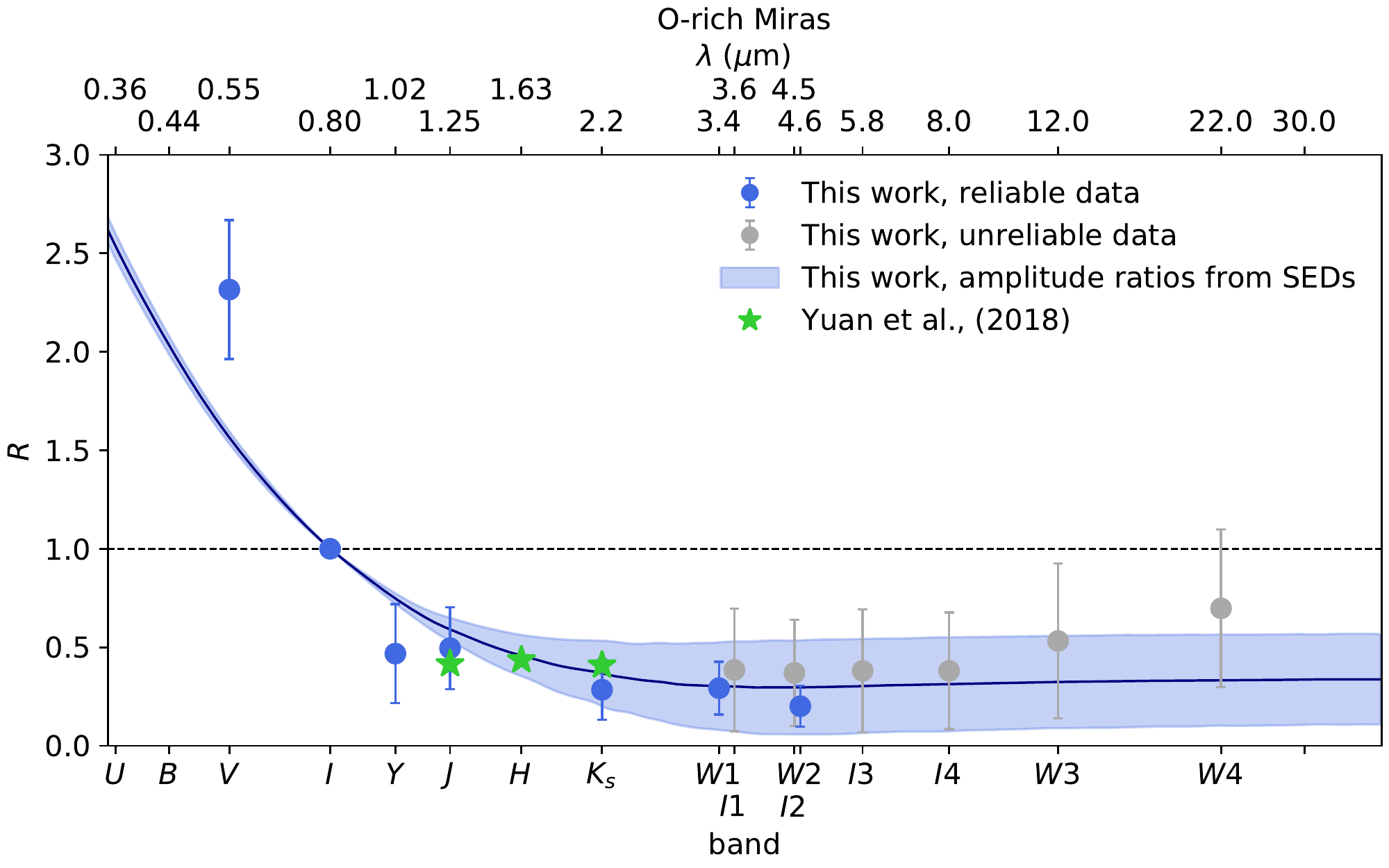} \\
\includegraphics[scale=0.6]{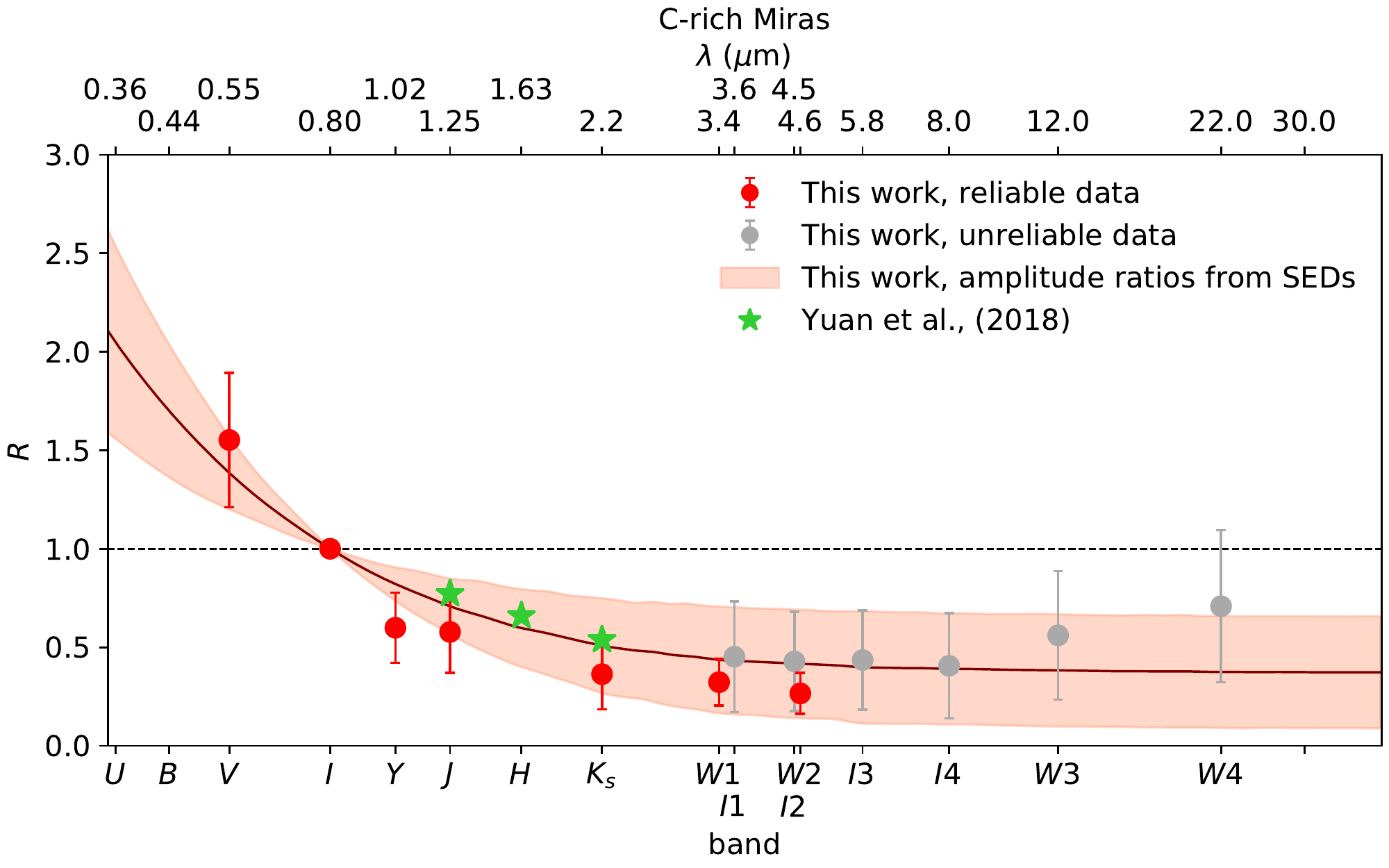}
\caption{Variability amplitude ratio $R$ as a function of wavelength for the O-rich (top panel) and C-rich (bottom panel) Miras (with $R=1$ fixed in $I$-band). The blue/red points represent the measurements, 
where the data sets had sufficient length and cadence to reliably measure the ratio, while the gray points represent light curves with typically two epochs leading to unreliable (degenerate) measurements of the ratio. The uncertainties of red, blue and gray points are calculated from the interquartile range (IQR)  as $1\sigma=0.741\times IQR$.
The green points are measurements taken from \cite{2018AJ....156..112Y}. The variability amplitude ratio $R$ decreases with the increasing wavelength as presented by the blue (top panel) and red (bottom panel) bands that are derived 
from the synthetic modeling of Miras as described in Section~\ref{sec:SED} (they are not best-fits to the data). The line inside the band is the mean value, while the band width reflects the bounds of calculated models.}
\label{fig:var_amp}
\end{figure*}

\begin{figure*}
\centering
\includegraphics[scale=0.6]{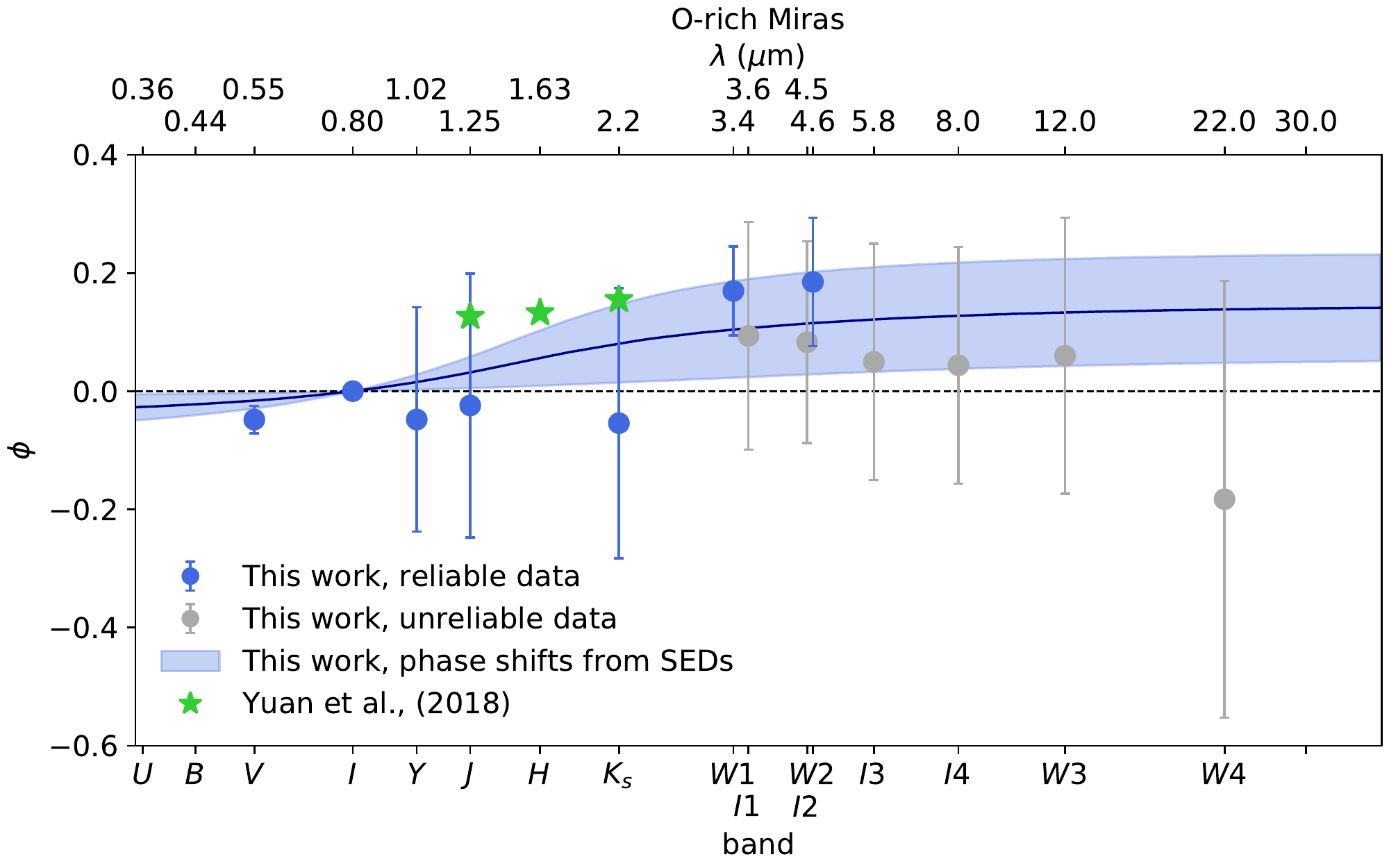} \\
\includegraphics[scale=0.6]{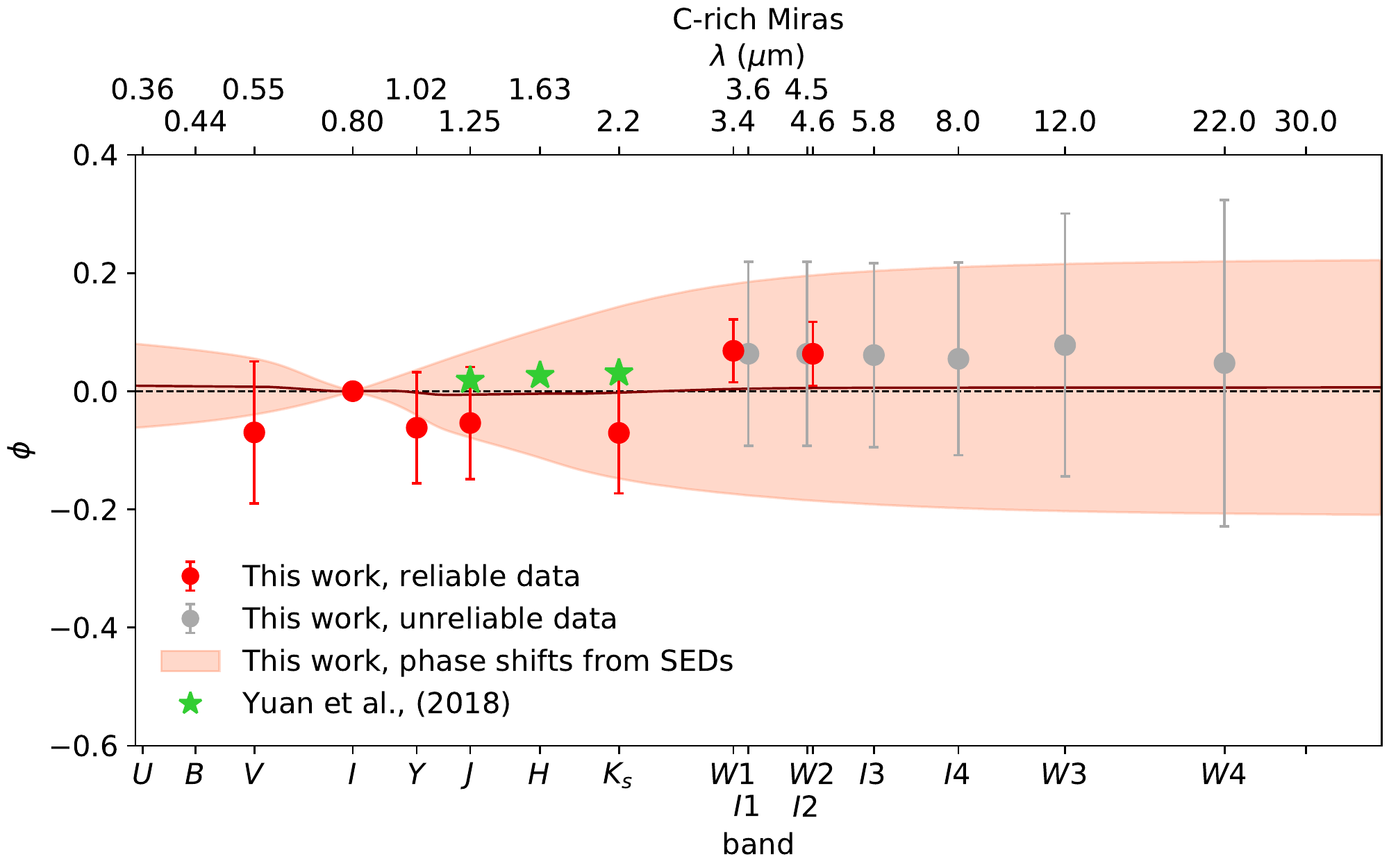}
\caption{Variability phase-lag $\phi$ as a function of wavelength for the O-rich (top panel) and C-rich (bottom panel) Miras are shown (with $\phi=0$ fixed in $I$-band). The blue/red points represent the measurements, 
where the data sets had sufficient length and cadence to reliably measure the phase-lag, while the gray points represent light curves with typically two epochs leading to unreliable (degenerate) measurements of the phase-lag. The uncertainties of red, blue and gray points are calculated from the interquartile range (IQR) as $1\sigma=0.741\times IQR$.
The green points are measurements taken from \cite{2018AJ....156..112Y}. The phase-lag $\phi$ slightly increases with the increasing wavelength as presented by the blue (top panel) and red (bottom panel) bands that are derived 
from the synthetic modeling of Miras as described in Section~\ref{sec:SED} (they are not best-fits to the data). The line inside the band is the mean value, while the band width reflects the bounds of calculated models.}
\label{fig:var_phase}
\end{figure*}

\begin{figure*}
\centering
\includegraphics[scale=0.6]{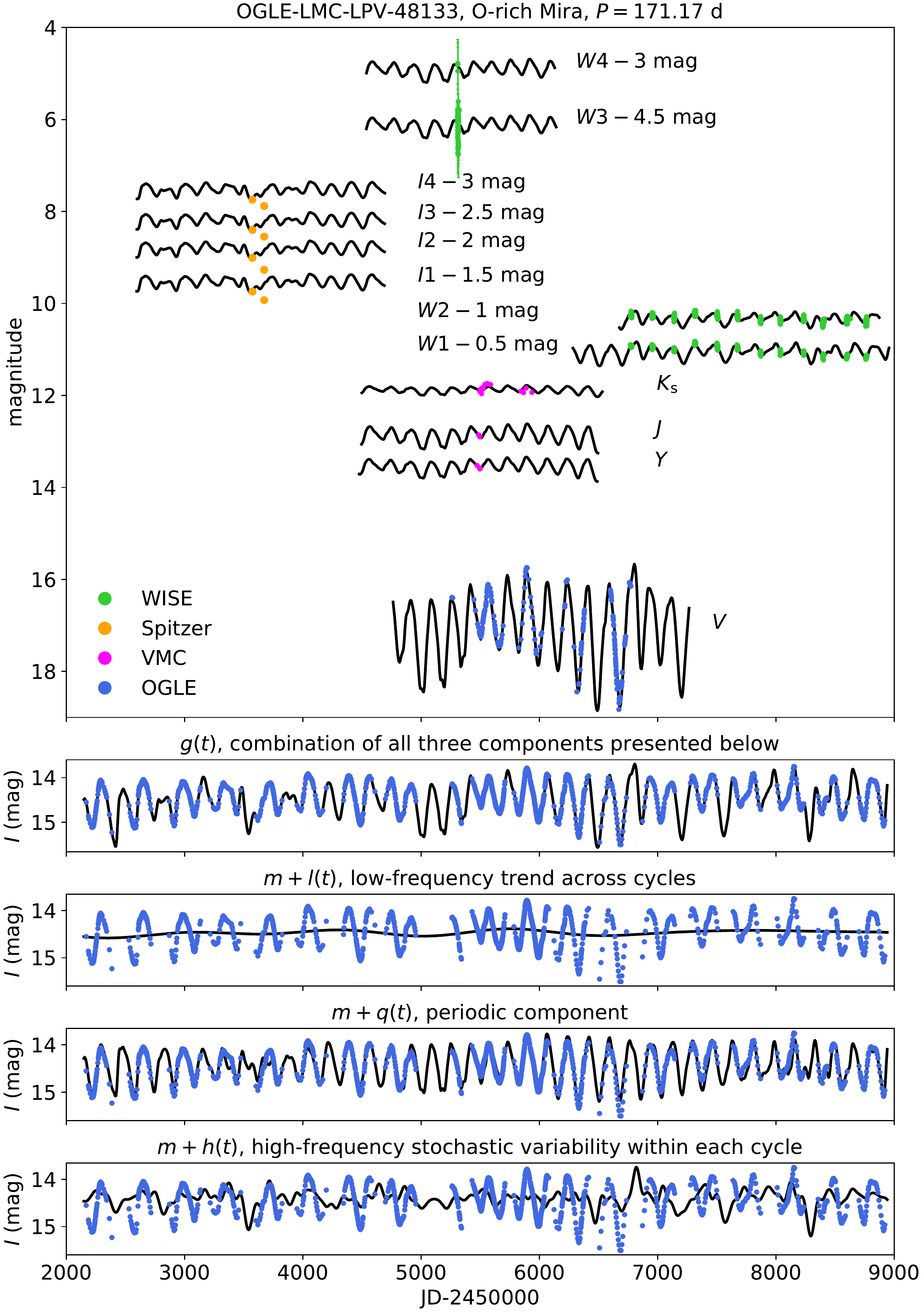}
\caption{Variability of an exemplary O-rich Mira in a number of optical--mid-IR wavelengths is shown. The Mira $I-$band light curve (blue points presented in the second panel) can be decomposed into three parts using model proposed by \citet{2016AJ....152..164H}. The solid black lines in the four bottom panels show the modeling of the OGLE $I-$band data (from the top: the combination of the three components $g(t)$ as in Equation \ref{eqn:model}, slowly-variable average magnitude component $l(t)$, periodic component $q(t)$, and stochastic  component $h(t)$). In the top panel, we show the phase-shifted, amplitude-scaled and magnitude-shifted $I$-band light curve model fitted to the 0.5--22 micron light curves for the considered surveys.}
\label{fig:lc_o}
\end{figure*}

\begin{figure*}
\centering
\includegraphics[scale=0.6]{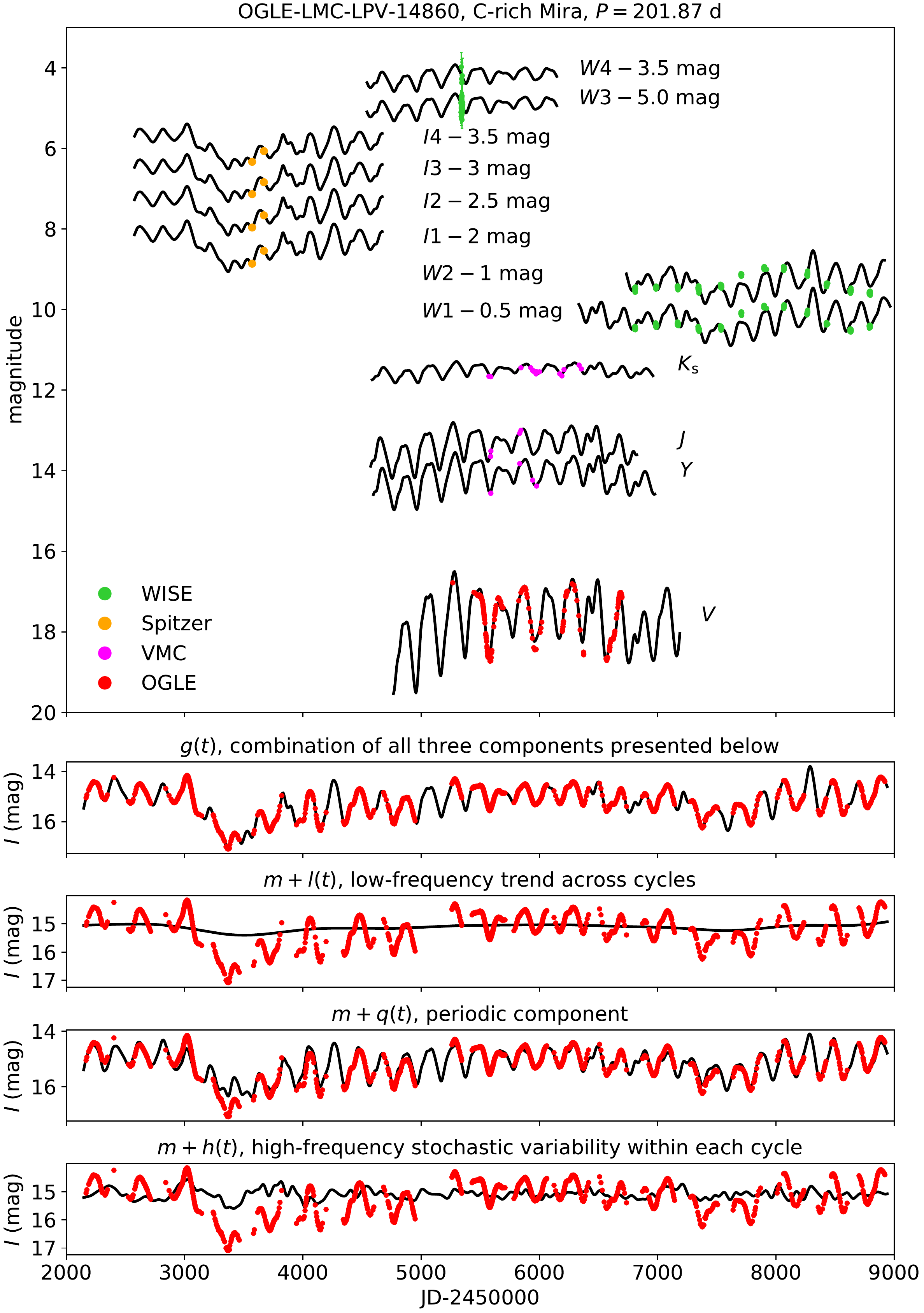}
\caption{Variability of an exemplary C-rich Mira in a number of optical--mid-IR wavelengths is shown. The Mira $I-$band light curve (red points presented in the second panel) can be decomposed into three parts using model proposed by \citet{2016AJ....152..164H}. The solid black lines in the four bottom panels show the modeling of the OGLE $I-$band data (from the top: the combination of the three components $g(t)$ as in Equation \ref{eqn:model}, slowly-variable average magnitude component $l(t)$, periodic component $q(t)$, and stochastic  component $h(t)$). In the top panel, we show the phase-shifted, amplitude-scaled and magnitude-shifted $I$-band light curve model fitted to the 0.5--22 micron light curves for the considered surveys.}
\label{fig:lc_c}
\end{figure*}

\subsection{Fitting templates to the NIR and mid-IR data}

Fitting the well-sampled $I$-band templates to the sparsely sampled near-IR and mid-IR data is key to a reliable calculation of the true mean magnitudes of Miras in these bands. The procedure turned out, however, to be non trivial due to insufficient sampling or data span of some of the data sets.

\begin{figure}
\includegraphics[scale=0.40]{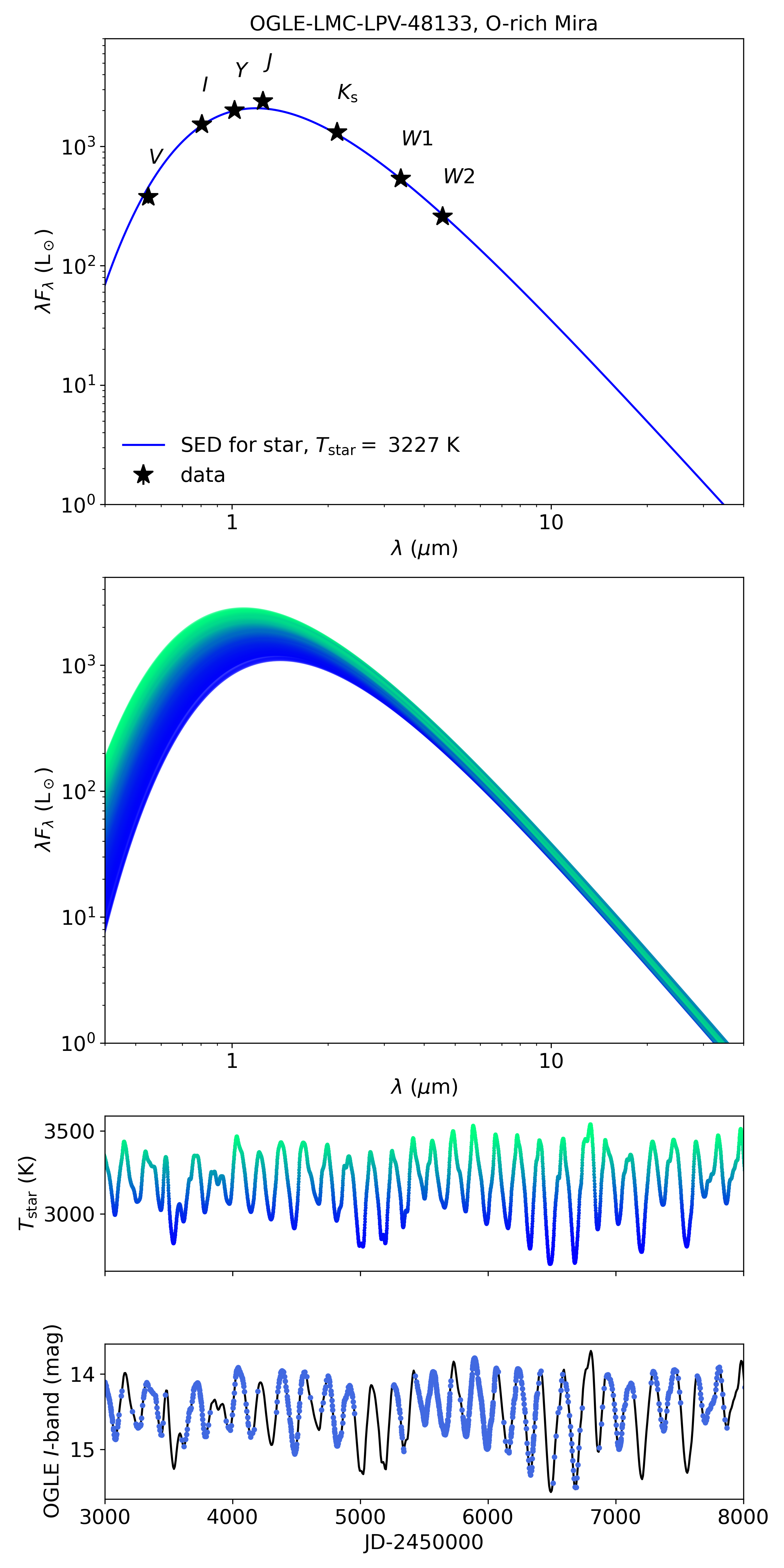}
\caption{SED for an O-rich Mira from the golden sample. The top panel shows the Planck function fit (blue line) to seven bands (stars). The mean temperature is $T_{\rm star}=3227~K$. The next panel shows a change in the SED shape with the color-coded SEDs with the colors from the third panel. In the third panel, we show the best-fit color-coded temperature as a function of time for the star, while in the bottom panel we show the OGLE $I$-band light curve (blue dots) along with the best-fit model (black solid line).}
\label{fig:SED_O}
\end{figure}

In the first method (method 1), we used a simple fitting of the $I$-band template to other bands by using a $\chi^2$ minimization method. We assumed that the shape of light curves is the same in each band, and the variability is a simple scaled, shifted in magnitude and shifted in phase version (hence three fitted parameters) of the $I$-band template. This methodology worked well for the $V$-, $YJK_\mathrm{s}$-, and $W1$ and $W2$-bands, as the data span and phase coverage were sufficient. The resulting variability amplitude ratios and phase-lags are presented in Figures~\ref{fig:var_amp} and \ref{fig:var_phase}, respectively, as blue points for the O-rich Miras and red points for the C-rich Miras. The uncertainties of both blue and red points presented in Figures~\ref{fig:var_amp} and \ref{fig:var_phase} are calculated from the interquartile range (IQR) as $1\sigma=0.741\times IQR$ for all obtained variability amplitude ratios and phase-lags in this method. We do not find correlations between the amplitude ratios or phase-lags and Mira pulsation periods in the mid-IR bands.

Because bands $I1$, $I2$, $I3$, $I4$, $W3$, and $W4$ contained typically two points only spanning a very narrow phase range, the resulting individual fits were unreliable (degenerate). We therefore tested a method, where we simultaneously fitted all light curves with their respectable models, all of them shifted by the same phase-lag and variability amplitude ratio, but individual magnitude shifts (method 2). The minimisation was over the sum of $\chi^2$ from individual fits. In the case of $M1$ observations, the fitting procedure is not possible due to the lack of observations epochs.

Another method (method 3) we used to study the sparse $I1$, $I2$, $I3$, $I4$, $W3$, and $W4$ bands was a method described in Section 6 of \citet{2005AcA....55..331S}. In short, each light curve consisted of two measurements. For a given pair of the variability amplitude ratio and phase-lag, we modified the template light curve and shifted it in magnitude so the first measurement ended up exactly on the modified template light curve and then again we modified the template light curve and shifted it in magnitude so the second measurement ended up exactly on the modified template light curve. We searched for a pair of the variability amplitude ratio/phase-lag parameters, where the difference between the two modified template light curves was the smallest.

\begin{figure}
\includegraphics[scale=0.4]{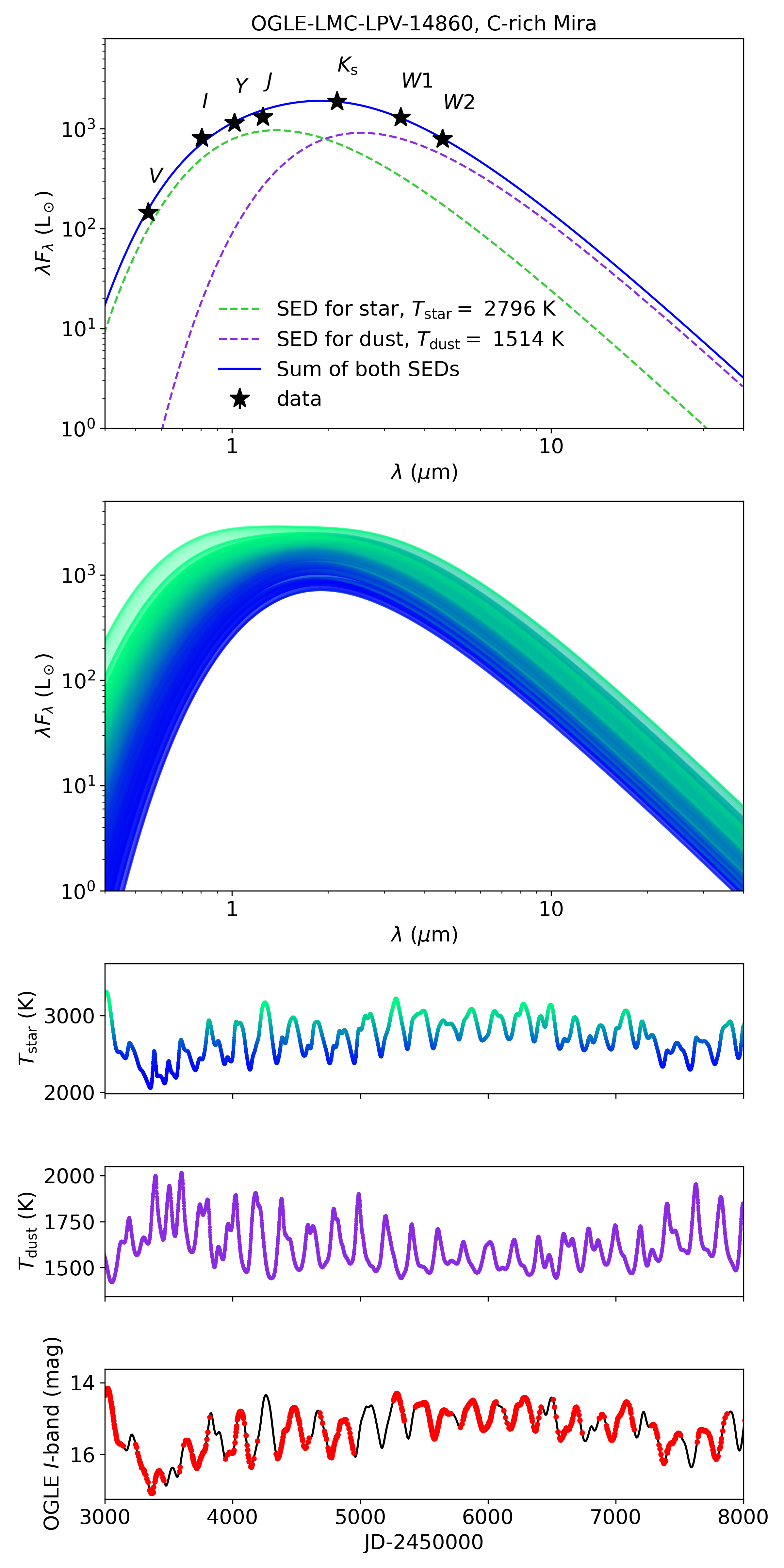}
\caption{SED for a C-rich Mira. The top panel shows the best fit (blue line) to seven bands (stars), where the dashed green line shows the Planck function for the star, the dashed violet line shows the Planck function for the dust, and the solid blue line is the combination of the two. The mean stellar temperature is $T_{\rm star}=2796~K$ and the mean dust temperature is $T_{\rm dust}=1514~K$. The next panel shows a change in the SED shape with the color-coded SEDs with the colors from the third panel. In the third panel, we show the best-fit color-coded star temperature as a function of time.
The fourth panel presents temperature of dust as a function of time, while in the bottom panel, we show the OGLE $I$-band light curve (red dots) along with the best-fit model (black solid line).}
\label{fig:SED_C}
\end{figure}

Light curves in the OGLE $VI$-bands, VMC $YJK_\mathrm{s}$-bands, Spitzer $I1$-, $I2$-, $I3$-, $I4$-bands, and WISE \mbox{$W1$-,} $W2$-, $W3$-, $W4$-bands are presented in Figures \ref{fig:lc_o} and \ref{fig:lc_c} for O- and C-rich Miras, respectively, along with $I-$band templates, template components and fitted templates to the near-IR and mid-IR data.

\section{The Observed variability amplitude ratio and phase-lag as a function of wavelength}
\label{sec:varpha}

In Figures~\ref{fig:var_amp} and \ref{fig:var_phase}, we present the variability amplitude ratio $R$ and phase-lag $\phi$ as a function of wavelength. 
These figures present both the real (dots/stars) and synthetic data (bands), the latter described in Section~\ref{subsec:synthetic_ratios}.

Let us concentrate on the blue (O-rich Miras) and red (C-rich Miras) points in these figures. Both the blue and red data points were calculated with the $\chi^2$ fitting (method 1) of the template $I$-band light curves to $V$, $Y$, $J$, $K_\mathrm{s}$, $W1$, and $W2$ light curves. Therefore, we explore the variability amplitude ratio $R$ and phase-lag $\phi$ with respect to the $I$-band light curves, where $R=1$ and $\phi=0$ are for the $I$-band. The negative (positive) $\phi$ means that the analysed light curve or band leads (lags) the $I$-band light curve.

From Figure~\ref{fig:var_amp}, it is clear that the variability amplitude ratio $R$ decreases with the increasing wavelength. For the O-rich Miras (top panel), the $V$-band variability amplitude is approximately 2.3 times greater than the amplitude in the $I$-band, while the $W1$ and $W2$ variability amplitude is approximately four times smaller than that in the $I$-band. We observe a similar dependence of the variability amplitude ratio for the C-rich Miras (bottom panel), albeit with a somewhat smaller amplitude (with approximately 1.6 greater variability in $V$-band) at short wavelengths as compared to the O-rich Miras.

The phase-lag picture is not as striking as the one for the variability amplitude ratio. From both panels in Figure~\ref{fig:var_phase}, a weak increase of the phase-lag $\phi$ with the increasing wavelength is noticeable. For both the O-rich (top panel) and C-rich (bottom panel) Miras the $V$-band leads the $I$-band, while at longer wavelengths it appears that a positive phase-lag is preferred, albeit with a rather choppy increase.

The grey points in Figures~\ref{fig:var_amp} and \ref{fig:var_phase} were derived with method 2 or 3, with the uncertainties calculated as $1\sigma=0.741\times IQR$. Due to a narrow time-span of the Spitzer and long-wavelength WISE data, effectively probing typically a small fraction of the phase, we conclude these measurements of the variability amplitude ratio $R$ and phase-lag $\phi$ are unreliable. They are presented here for completeness.

\section{Synthetic properties of Miras}
\label{sec:SED}

The magnitudes of a star observed in multiple filters may be converted to ``average in-band'' (or ``in-filter'') fluxes $\lambda F_\lambda$ (in units of erg/s or $L_\odot$), where $F_\lambda$ is the spectral flux density, being the flux per unit wavelength. To form the SED of that star, the zero-magnitude fluxes and transmission properties of filters must be known. To model such an observed or synthetic SED one may use the Planck function(s) that need to be converted into in-band average fluxes $F=\int W(\lambda) \lambda F_\lambda d\lambda$ by using the filter transmissions $W(\lambda)$. To find a best-fitting SED model, we used a standard $\chi^2$ minimization procedure, where the model parameters were either two or four parameters: the amplitude(s) and temperature(s) of the Planck function(s). The median uncertainty of the stellar temperature is 6\%, while for the dust it is 10\%.

Detailed analyses of SEDs showed that mean magnitudes in both $W3$- and $W4$-bands were not reliable for many individual Miras. The $W3$- and especially $W4$-band measurements appeared as outliers in the SED models, while the $M1$-band (24 microns) measurements seemed to fit rather well in many cases (measurements in $W4$ at 22 microns and $M1$ at 24 microns often disagreed by as much as an order of magnitude in luminosity). 

In our sample of 1663 Miras, there was a sub-sample of \goldnum sources ($29$ O-rich and $111$ C-rich Miras, hereafter the golden sample) with complete and high quality $V$, $I$, $Y$, $J$, $K_\mathrm{s}$, $W1$, and $W2$ light curves, for which we simultaneously measured the variability amplitude ratios, phase lags, and magnitude shifts very precisely (with method 1). Having the derived model parameters, we were able to shift and scale the $I$-band templates to the remaining bands.

Equipped with a complete information about their extinction corrected brightness in the $V$, $I$, $Y$, $J$, $K_\mathrm{s}$, $W1$, and $W2$ bands for the duration of multiple periods and with hundreds of model epochs per period, we converted these magnitudes to fluxes ($\lambda F_\lambda$ in $L_\odot$) assuming the LMC distance of $d_{\rm LMC}=49.59$~kpc (\citealt{2019Natur.567..200P}) and the respective filter parameters/transmissions. Then, for each star, we created 6000 synthetic SEDs spanning 6000 days (one SED per day). We modeled these SEDs with either a single Planck (Figure~\ref{fig:SED_O}) or double Planck (Figure~\ref{fig:SED_C}) functions, where the double Planck function means a sum of two Planck functions: the hotter one for the star and the cooler one (most likely) for the dust. SEDs for all considered O-rich Miras did not require contribution from the dust, while such contribution was frequently necessary for the C-rich Miras.

\subsection{Variability amplitude ratio and phase-lag as a function of wavelength} \label{subsec:synthetic_ratios}

Using 6000 SEDs per star, we created synthetic light curves between 0.1--40 $\mu$m and spaced every 0.1 $\mu$m. Each synthetic light curve was then checked against the 0.8 $\mu$m one (assumed to represent $I$-band) for the shift in time and the variability amplitude ratio. For each spacing in wavelength, we obtained \goldnum variability amplitude ratios and phase shifts. The solid continuous area in Figures~\ref{fig:var_amp} and \ref{fig:var_phase} represents the range of these synthetic variability amplitude ratios and/or phase-lags. The solid line in the middle of the shaded band is the mean value. These color bands are not fits to the data. Both the variability amplitude ratios and phase-lags are also presented in Table \ref{tab:amp_ratio} and \ref{tab:PhaseLag}, respectively, for the O- and C-rich Miras, along with spread of calculated models.

\vspace{0.5cm}
\begin{table}[]
\caption{Variability amplitude ratio $R$ as a function of wavelength for the O- and C-rich Miras.\label{tab:amp_ratio}}
\vspace{0.2cm}
\begin{center}
\begin{tabular}{l|c|c}
\hline
\multirow{2}{*}{$\lambda$ ($\mu$m)} & \multicolumn{2}{c}{$R$} \\ \cline{2-3}
 &  O-rich Miras & C-rich Miras \\ \hline \hline
0.1 & 6.863 $\pm$ 0.353 & 5.768 $\pm$ 1.473 \\
0.2 & 4.718 $\pm$ 0.098 & 3.555 $\pm$ 1.160 \\
0.3 & 3.115 $\pm$ 0.075 & 2.438 $\pm$ 0.678 \\
0.4 & 2.260 $\pm$ 0.055 & 1.855 $\pm$ 0.405 \\
0.5 & 1.748 $\pm$ 0.037 & 1.510 $\pm$ 0.250 \\
0.6 & 1.413 $\pm$ 0.027 & 1.280 $\pm$ 0.130 \\
0.7 & 1.175 $\pm$ 0.015 & 1.123 $\pm$ 0.062 \\
0.8 & 1.000 $\pm$ 0.000 & 1.000 $\pm$ 0.000 \\
0.9 & 0.873 $\pm$ 0.013 & 0.905 $\pm$ 0.045 \\
1.0 & 0.765 $\pm$ 0.025 & 0.833 $\pm$ 0.078 \\
$\vdots$ & $\vdots$ & $\vdots$ \\
40.0 & 0.338 $\pm$ 0.233 & 0.373 $\pm$ 0.283\\
\hline \hline
\end{tabular}
\end{center}
\tablecomments{The variability amplitude ratio $R$ is defined as a ratio between the amplitude at wavelength $\lambda$ and the amplitude in $I$-band. Values presented here represent the mean value of the variability amplitude ratios obtained from the synthetic light curves (marked by solid lines in Figure \ref{fig:var_amp}), while the band reflects the spread of calculated models. This table is available in its entirety in a machine-readable form in the online journal. A portion is shown here for guidance regarding its form and content.}
\end{table}

\vspace{0.5cm}
\begin{table}[]
\caption{Phase-lag $\phi$ as a function of wavelength for the O- and C-rich Miras.\label{tab:PhaseLag}}
\vspace{0.2cm}
\begin{center}
\begin{tabular}{l|c|c}
\hline
\multirow{2}{*}{$\lambda$ [$\mu$m]} & \multicolumn{2}{c}{$\phi$} \\ \cline{2-3}
 &  O-rich Miras & C-rich Miras \\ \hline \hline
0.1 & -0.03926 $\pm$ 0.03154 & 0.01711 $\pm$ 0.09838 \\
0.2 & -0.03480 $\pm$ 0.02818 & 0.01340 $\pm$ 0.08774 \\
0.3 & -0.02989 $\pm$ 0.02437 & 0.00999 $\pm$ 0.07646 \\
0.4 & -0.02451 $\pm$ 0.02009 & 0.00868 $\pm$ 0.06570 \\
0.5 & -0.01890 $\pm$ 0.01558 & 0.00816 $\pm$ 0.05415 \\
0.6 & -0.01306 $\pm$ 0.01085 & 0.00661 $\pm$ 0.03906 \\
0.7 & -0.00653 $\pm$ 0.00542 & 0.00128 $\pm$ 0.01797 \\
0.8 & 0.00000 $\pm$ 0.00000 & 0.00000 $\pm$ 0.00000 \\
0.9 & 0.00699 $\pm$ 0.00588 & 0.00100 $\pm$ 0.01601 \\
1.0 & 0.01398 $\pm$ 0.01177 & -0.00141 $\pm$ 0.03448 \\
$\vdots$ & $\vdots$ & $\vdots$ \\
40.0 & 0.14125 $\pm$ 0.09047 & 0.00632 $\pm$ 0.21544\\ \hline \hline
\end{tabular}
\end{center}
\tablecomments{The phase-lag $\phi$ is defined as a shift between a light curve at wavelength $\lambda$ and the $I$-band light curve. Values presented here represent the mean value of phase-lags obtained from the synthetic light curves (marked by solid lines in Figure \ref{fig:var_phase}), while the band reflects the spread of calculated models. This table is available in its entirety in a machine-readable form in the online journal. A portion is shown here for guidance regarding its form and content.}
\end{table}

\subsection{Bolometric luminosities} \label{sec:bolo}

From fitting a single or a double Planck function to the SED, we obtained both the amplitude and temperature of each Planck function. It was then straightforward to calculate the total bolometric luminosity for each Mira, as well as the separate bolometric luminosities for the star and dust. We integrated the synthetic SEDs in the 0--100 $\mu$m range. In Table~\ref{tab:bolo}, we provide the bolometric luminosity for both the star and dust, and the combined luminosity.

None of the O-rich Miras required contribution from the dust to the SED, while it was frequently necessary in the C-rich stars. The O-rich Miras have the bolomeric luminosity in a range of 3600--29200~$L_\odot$ with the median value of 13500~$L_\odot$. The bolometric luminosity for the C-rich Miras span a range of 800--50400~$L_\odot$ with the median value of 19200~$L_\odot$. The second (dust) component is present in 73\% of the C-rich Miras, where the bolometric luminosity of the dust spans a range of 300--42100~$L_\odot$ with the median value of 12500~$L_\odot$. We do not find a correlation between the bolometric luminosity and the temperature.

\begin{table*}[ht!]
\caption{Bolometric luminosities of Miras from the golden sample.}
\label{tab:bolo}
\vspace{0.2cm}
\begin{tabular}{l|c|c|c|c|c|c}
\hline \hline
ID & type & $L_{\rm bol}$ ($L_\odot$) & $L_{\rm bol}^{\rm star}$ ($L_\odot$) & $L_{\rm bol}^{\rm dust}$ ($L_\odot$) & $f^{\rm star}$ & $f^{\rm dust}$ \\
\hline \hline
OGLE-LMC-LPV-08058 & C & $32139 \pm 752$  & 9882 & 22257 & 0.307 & 0.693 \\
OGLE-LMC-LPV-08424 & C & $46070 \pm 1837$  & 3934 & 42136 & 0.085 & 0.915 \\
OGLE-LMC-LPV-09268 & C & $20767 \pm 1859$ & 17849  & 2918 & 0.860 & 0.140 \\
OGLE-LMC-LPV-10812 & C & $18032 \pm 446$  & 3656 & 14376 & 0.203 & 0.797 \\
OGLE-LMC-LPV-12829 & C & $24235\pm 1051$ & 24235   &   0 & 1.000 & 0.000 \\
OGLE-LMC-LPV-13563 & C & $20241 \pm 960$  & 9696 & 10545 & 0.479 & 0.521 \\
OGLE-LMC-LPV-13815 & C & $15053\pm  383$  & 3102 & 11951 & 0.206 & 0.794 \\
OGLE-LMC-LPV-14860 & C & $15281 \pm 508$  & 7882  & 7399 & 0.516 & 0.484 \\
OGLE-LMC-LPV-15353 & C & $14580 \pm 802$ & 10564  & 4016 & 0.725 & 0.275 \\
OGLE-LMC-LPV-16684 & C  & $1904 \pm  32$   & 522  & 1382 & 0.274 & 0.726 \\
\vdots & \vdots & \vdots & \vdots & \vdots & \vdots & \vdots \\
OGLE-LMC-LPV-82526 & O  & $3648 \pm  244$  & 3648     & 0 & 1.000 & 0.000 \\
\hline \hline
\end{tabular}
\tablecomments{The reported star ID and the O/C type  is from the catalog of \citet{2009AcA....59..239S}. 
Table rows are sorted by the star ID. $L_{\rm bol}$ is the total bolometric luminosity presented with its uncertainty, $L_{\rm bol}^{\rm star}$ and $L_{\rm bol}^{\rm dust}$ are the bolometric luminosity of the star and dust, while $f^{\rm star}=L_{\rm bol}^{\rm star}/L_{\rm bol}$ and $f^{\rm dust}=L_{\rm bol}^{\rm dust}/L_{\rm bol}$ are fractions of the bolometric luminosity of the star and dust to the total bolometric luminosity (and $f^{\rm star} + f^{\rm dust} =1$).
This table is available in its entirety in a machine-readable form in the online journal. A portion is shown here for guidance regarding its form and content.}
\end{table*}
\vspace{3mm}



\subsection{Synthetic PLRs}

\vspace{0.5cm}
\begin{table}[]
\caption{Synthetic PLR slope $a_1$ derived from the SED analysis, as a function of wavelength, for the O- and C-rich Miras.
\label{tab:synthetic_slope}}
\vspace{0.2cm}
\begin{tabular}{l|c|c}
\hline
\multirow{2}{*}{$\lambda$ [$\mu$m]} & \multicolumn{2}{c}{$a_1$} \\ \cline{2-3}
                        & O-rich Miras     & C-rich Miras    \\
\hline \hline
$0.1$ & $42.892 \pm 13.814$ & $47.341 \pm 11.189$ \\
$0.2$ & $18.591 \pm 6.690$ & $23.555 \pm 5.113$ \\
$0.3$ & $10.490 \pm 4.320$ & $15.627 \pm 3.129$ \\
$0.4$ & $6.440 \pm 3.138$ & $11.663 \pm 2.177$ \\
$0.5$ & $4.010 \pm 2.431$ & $9.288 \pm 1.647$ \\
$0.6$ & $2.390 \pm 1.964$ & $7.714 \pm 1.332$ \\
$0.7$ & $1.235 \pm 1.632$ & $6.602 \pm 1.144$ \\
$0.8$ & $0.370 \pm 1.387$ & $5.774 \pm 1.032$ \\
$0.9$ & $-0.301 \pm 1.199$ & $5.114 \pm 0.963$ \\
$1.0$ & $-0.835 \pm 1.052$ & $4.534 \pm 0.914$ \\
$\vdots$ & $\vdots$ & $\vdots$ \\
$40.0$ & $-4.529 \pm 0.419$ & $-9.607 \pm 0.586$ \\
\hline \hline
\end{tabular}
\tablecomments{The PLR slope $a_1$ from the linear fit in a form $m_\lambda = a_0 + a_1 \times (\log P-2.3)$. This table is available in its entirety in a machine-readable form in the online journal. A portion is shown here for guidance regarding its form and content.}
\end{table}

Equipped with \goldnum Miras in the selected golden sample, each with a mean SED as well as a series of 6000 synthetic SEDs as a function of time, we investigated the dependence of the PLR slope on the wavelength, but also the temperature and color changes as a function of time.

\begin{figure}
\centering
\includegraphics[scale=0.35]{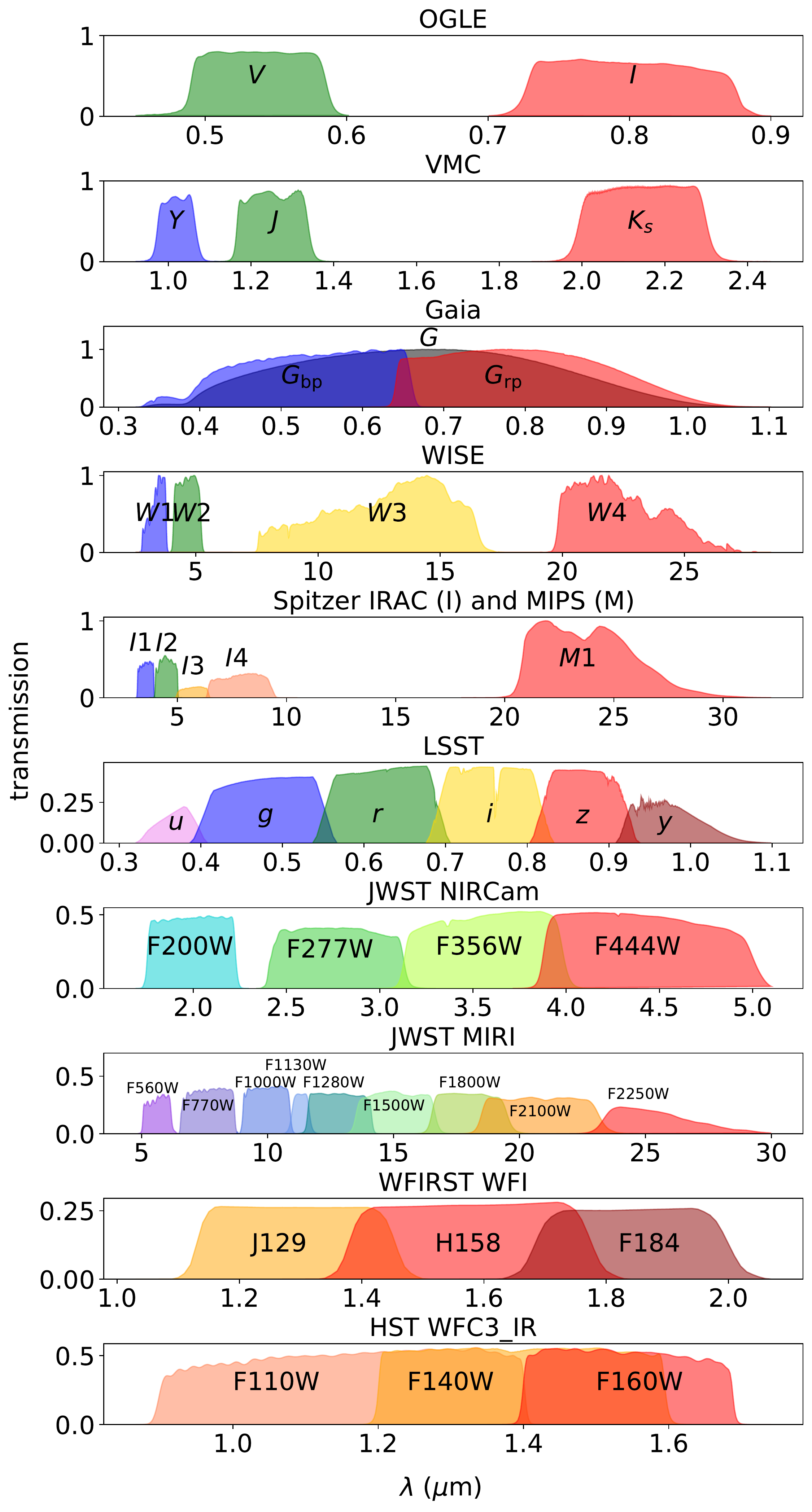}
\caption{Transmission curves for filters used in this paper. In our analyses, we included bands from the existing and working surveys, such as OGLE, VMC, Gaia, WISE, Spitzer, HST, but we also considered filters from future facilities, such as The Vera C. Rubin Observatory (formerly LSST), JWST, and The Nancy Grace Roman Space Telescope (formerly WFIRST).}
\label{fig:filters}
\end{figure}

We used these SEDs to create 42 synthetic PLRs in frequently used filters in major present/future surveys. They include 
$V$, $I$ from OGLE,
$Y$, $J$, $K_\mathrm{s}$ from VMC,
$G$, $G_{bp}$, $G_{rp}$ from Gaia,
$u$, $g$, $r$, $i$, $z$, $y$ from The Vera C. Rubin Observatory,
$F110W$, $F140W$, $F160W$ from HST,
$J129$, $H158$, $F184$ from The Nancy Grace Roman Space Telescope,
$F200W$, $F277W$, $F356W$, $F444W$, $F560$, $F770W$, $F1000$, $F1130W$, $F1280$, $F1500W$, $F1800W$, $F2100W$, $F2550W$ from JWST,
$W1$, $W2$, $W3$, $W4$ (3.4, 4.6, 12, 22 micron) from WISE, and
$I1$, $I2$, $I3$, $I4$, $M1$ (3.6, 4.5, 5.8, 8.0, 24 micron) from IRAC/MIPS Spitzer.
The transmission curves for all mentioned bands are presented in Figure \ref{fig:filters}.
Magnitudes in all filters are provided in the Vega magnitude system with the exception of the LSST filters, where the magnitudes are provided in the AB magnitude system.

\begin{figure*}
\centering
\includegraphics[scale=0.26]{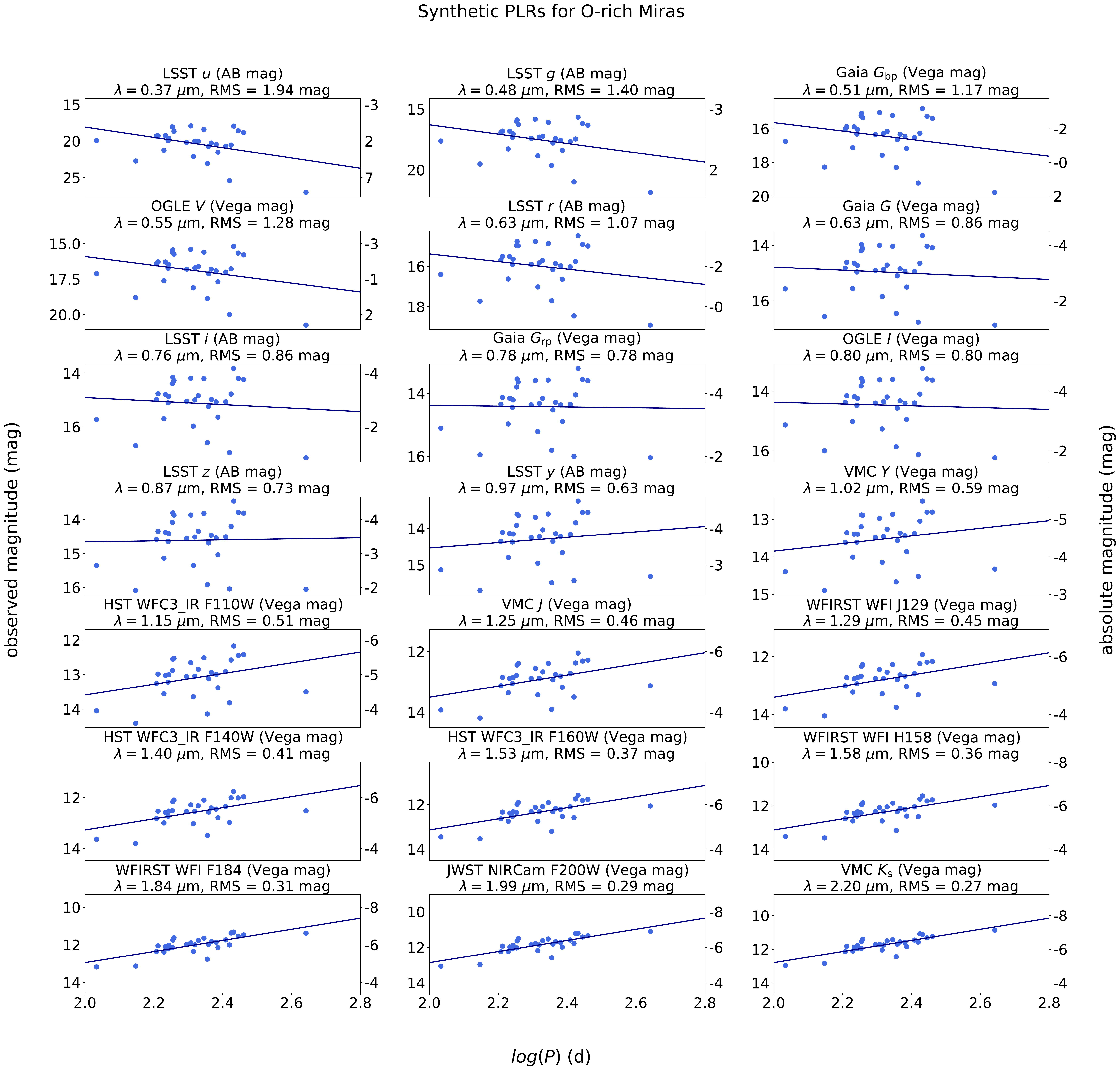} \\
\caption{Synthetic PLRs for the golden sample O-rich Miras in the 0.37--2.20 micron range. Each panel is labeled with a survey filter and the wavelength. The left (right) y-axis shows the observed (absolute) magnitude. The absolute magnitude is calculated using distance modulus $\mu = 18.477$ mag \citep{2019Natur.567..200P}.}
\label{fig:PLR_O1}
\end{figure*}

\begin{figure*}
\centering
\includegraphics[scale=0.26]{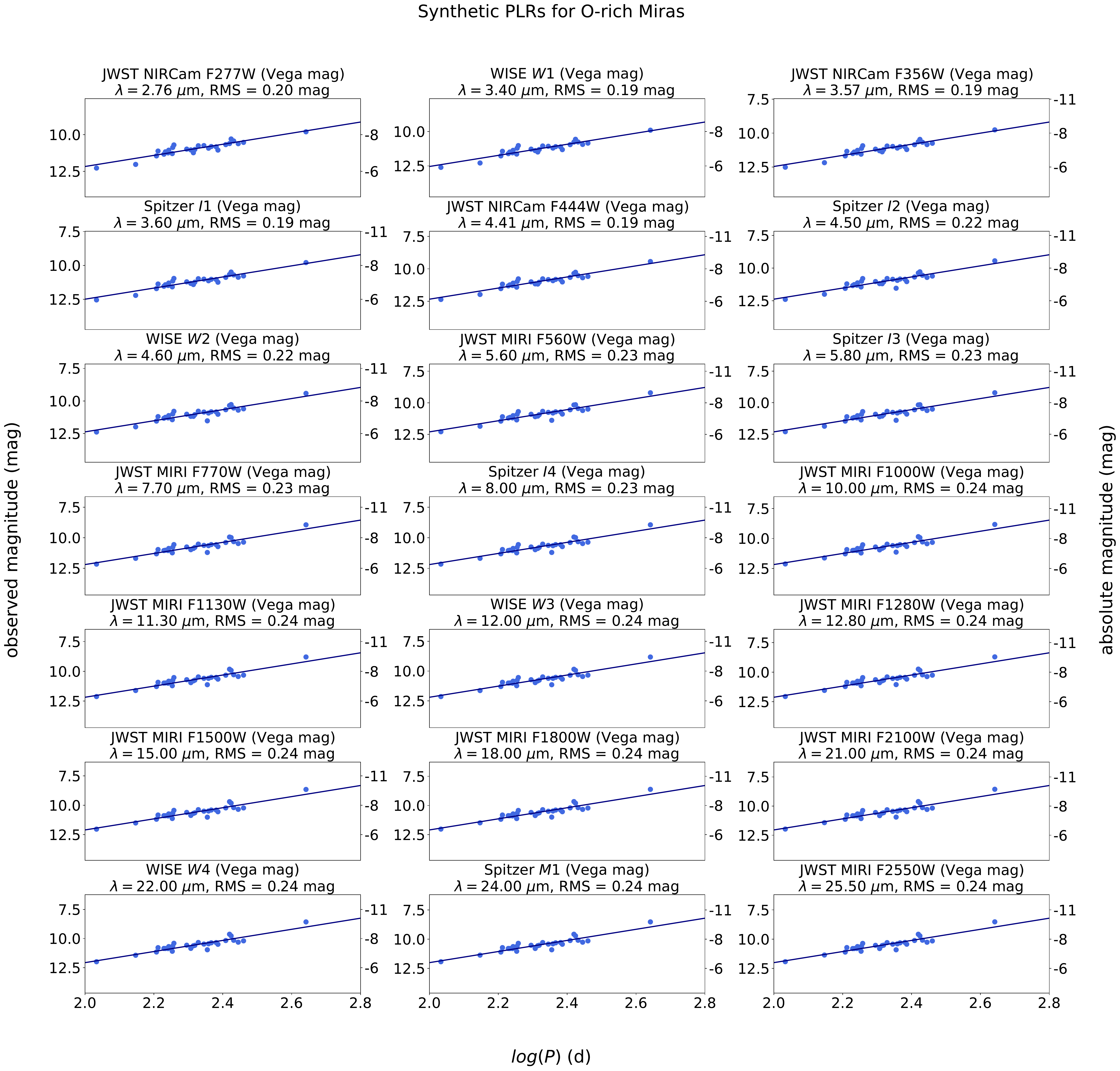} \\
\caption{Synthetic PLRs for the golden sample O-rich Miras in the 2.76--25.50 micron range. Each panel is labeled with a survey filter and the wavelength. The left (right) y-axis shows the observed (absolute) magnitude. The absolute magnitude is calculated using distance modulus $\mu = 18.477$ mag \citep{2019Natur.567..200P}.}
\label{fig:PLR_O2}
\end{figure*}

\begin{figure*}
\centering
\includegraphics[scale=0.26]{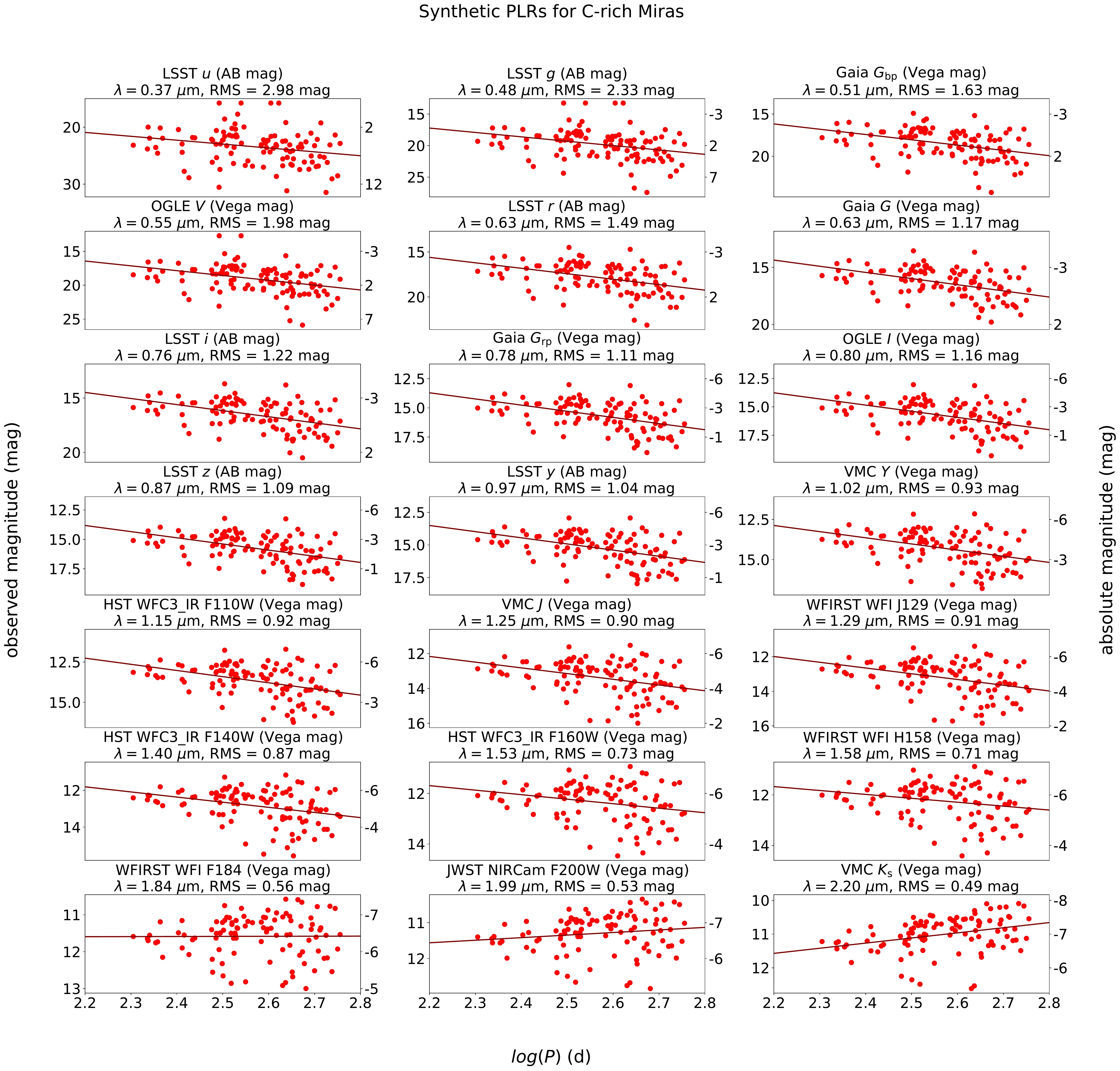} \\
\caption{Synthetic PLRs for the golden sample C-rich Miras in the 0.37--2.20 micron range. Each panel is labeled with a survey filter and the wavelength. The left (right) y-axis shows the observed (absolute) magnitude. The absolute magnitude is calculated using distance modulus $\mu = 18.477$ mag \citep{2019Natur.567..200P}.}
\label{fig:PLR_C1}
\end{figure*}

\begin{figure*}
\centering
\includegraphics[scale=0.26]{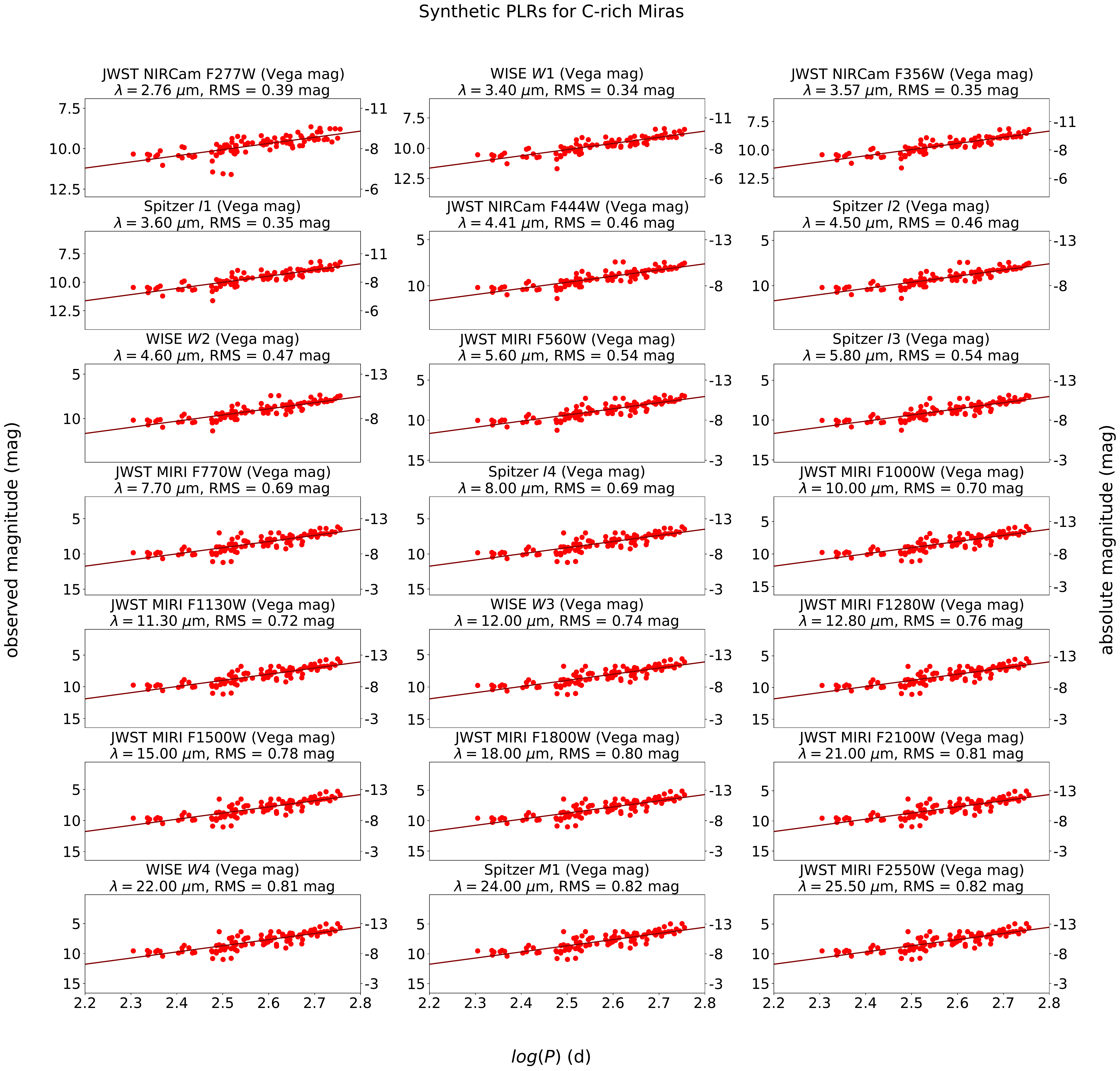} \\
\caption{Synthetic PLRs for the golden sample C-rich Miras in the 2.76--25.50 micron range. Each panel is labeled with a survey filter and the wavelength. The left (right) y-axis shows the observed (absolute) magnitude. The absolute magnitude is calculated using distance modulus $\mu = 18.477$ mag \citep{2019Natur.567..200P}.}
\label{fig:PLR_C2}
\end{figure*}

For each star, we calculated its synthetic, calibrated (extinction-free) magnitudes in all 42 bands. We then fitted a linear relation in a form $m_\lambda = a_0 + a_1 \times (\log P-2.3)$ in each filter to either O-rich (in the range $2.0 < \log P < 2.65 $) or C-rich (in the range $2.3 < \log P < 2.76 $) Miras. During the fitting procedure, we applied the $\sigma$-clipping procedure rejecting outliers deviating more than $3\sigma$ from the fit. The synthetic magnitudes as well as the best-fit PLRs are presented in Figures~\ref{fig:PLR_O1}, \ref{fig:PLR_O2}, \ref{fig:PLR_C1}, and \ref{fig:PLR_C2}. From these figures it is clear that the PLR slope changes from having the positive sign in $u$-band to having the negative sign at near- and mid-IR wavelengths. In Tables~\ref{tab:PLRparO} and \ref{tab:PLRparC}, we present the model parameters for the synthetic PLRs.

\begin{table*}[ht!]
\caption{Parameters of synthetic PLRs for the O-rich Miras. }
\label{tab:PLRparO}
\vspace{0.2cm}
\footnotesize
\begin{tabular}{l|r|ccc|c}
\hline
\multicolumn{5}{c}{O-rich Miras} \\
\hline \hline
Survey and filter names & $\lambda_{\mathrm{eff}}$ ($\mu$m) & $a_{\mathrm{0, obs}}$ & $a_{\mathrm{0, abs}}$ & $a_1$ & RMS (mag)\\ 
\hline \hline
LSST $u$ (AB mag) & $0.37$ & $20.190 \pm 0.380$ & $1.713 \pm 0.381$ & $7.064 \pm 3.308$ & $1.94$ \\
 LSST $g$ (AB mag) & $0.48$ & $17.442 \pm 0.273$ & $-1.035 \pm 0.274$ & $3.830 \pm 2.378$ & $1.40$ \\ 
 Gaia $G_{\mathrm{bp}}$ (Vega mag) & $0.51$ & $16.383 \pm 0.230$ & $-2.094 \pm 0.231$ & $2.491 \pm 1.998$ & $1.17$ \\ 
 OGLE $V$ (Vega mag) & $0.55$ & $16.841 \pm 0.249$ & $-1.636 \pm 0.251$ & $3.126 \pm 2.169$ & $1.28$ \\ 
 LSST $r$ (AB mag) & $0.63$ & $15.947 \pm 0.209$ & $-2.530 \pm 0.210$ & $1.892 \pm 1.816$ & $1.07$ \\ 
 Gaia $G$ (Vega mag) & $0.63$ & $14.949 \pm 0.167$ & $-3.528 \pm 0.169$ & $0.557 \pm 1.456$ & $0.86$ \\ 
 LSST $i$ (AB mag) & $0.76$ & $15.105 \pm 0.168$ & $-3.372 \pm 0.170$ & $0.653 \pm 1.462$ & $0.86$ \\ 
 Gaia $G_{\mathrm{rp}}$ (Vega mag) & $0.78$ & $14.412 \pm 0.152$ & $-4.065 \pm 0.154$ & $0.128 \pm 1.323$ & $0.78$ \\ 
 OGLE $I$ (Vega mag) & $0.80$ & $14.457 \pm 0.157$ & $-4.020 \pm 0.159$ & $0.301 \pm 1.364$ & $0.80$ \\ 
 LSST $z$ (AB mag) & $0.87$ & $14.611 \pm 0.142$ & $-3.866 \pm 0.144$ & $-0.151 \pm 1.236$ & $0.73$ \\ 
 LSST $y$ (AB mag) & $0.97$ & $14.311 \pm 0.124$ & $-4.166 \pm 0.126$ & $-0.737 \pm 1.075$ & $0.63$ \\ 
 VMC $Y$ (Vega mag) & $1.02$ & $13.544 \pm 0.115$ & $-4.933 \pm 0.118$ & $-1.016 \pm 1.003$ & $0.59$ \\ 
 HST WFC3\_IR F110W (Vega mag) & $1.15$ & $13.124 \pm 0.099$ & $-5.353 \pm 0.102$ & $-1.549 \pm 0.864$ & $0.51$ \\ 
 VMC $J$ (Vega mag) & $1.25$ & $12.957 \pm 0.090$ & $-5.520 \pm 0.094$ & $-1.832 \pm 0.786$ & $0.46$ \\ 
 WFIRST WFI J129 (Vega mag) & $1.29$ & $12.826 \pm 0.088$ & $-5.651 \pm 0.091$ & $-1.922 \pm 0.764$ & $0.45$ \\ 
 HST WFC3\_IR F140W (Vega mag) & $1.40$ & $12.617 \pm 0.080$ & $-5.860 \pm 0.084$ & $-2.182 \pm 0.700$ & $0.41$ \\ 
 HST WFC3\_IR F160W (Vega mag) & $1.53$ & $12.393 \pm 0.072$ & $-6.084 \pm 0.076$ & $-2.487 \pm 0.626$ & $0.37$ \\ 
 WFIRST WFI H158 (Vega mag) & $1.58$ & $12.343 \pm 0.070$ & $-6.134 \pm 0.074$ & $-2.555 \pm 0.611$ & $0.36$ \\ 
 WFIRST WFI F184 (Vega mag) & $1.84$ & $12.060 \pm 0.060$ & $-6.417 \pm 0.065$ & $-2.963 \pm 0.524$ & $0.31$ \\ 
 JWST NIRCam F200W (Vega mag) & $1.99$ & $11.925 \pm 0.057$ & $-6.552 \pm 0.061$ & $-3.131 \pm 0.492$ & $0.29$ \\ 
 VMC $K_\mathrm{s}$ (Vega mag) & $2.20$ & $11.804 \pm 0.053$ & $-6.673 \pm 0.058$ & $-3.295 \pm 0.463$ & $0.27$ \\ 
 JWST NIRCam F277W (Vega mag) & $2.76$ & $11.035 \pm 0.039$ & $-7.442 \pm 0.046$ & $-3.785 \pm 0.335$ & $0.20$ \\ 
 WISE $W1$ (Vega mag) & $3.40$ & $11.328 \pm 0.037$ & $-7.149 \pm 0.044$ & $-4.040 \pm 0.320$ & $0.19$ \\ 
 JWST NIRCam F356W (Vega mag) & $3.57$ & $11.234 \pm 0.037$ & $-7.243 \pm 0.044$ & $-4.099 \pm 0.318$ & $0.19$ \\ 
 Spitzer $I1$ (Vega mag) & $3.60$ & $11.262 \pm 0.037$ & $-7.215 \pm 0.044$ & $-4.097 \pm 0.318$ & $0.19$ \\ 
 JWST NIRCam F444W (Vega mag) & $4.41$ & $11.056 \pm 0.037$ & $-7.421 \pm 0.044$ & $-4.288 \pm 0.321$ & $0.19$ \\ 
 Spitzer $I2$ (Vega mag) & $4.50$ & $11.102 \pm 0.044$ & $-7.375 \pm 0.050$ & $-4.242 \pm 0.382$ & $0.22$ \\ 
 WISE $W2$ (Vega mag) & $4.60$ & $11.094 \pm 0.044$ & $-7.383 \pm 0.050$ & $-4.264 \pm 0.382$ & $0.22$ \\ 
 JWST MIRI F560W (Vega mag) & $5.60$ & $10.980 \pm 0.044$ & $-7.497 \pm 0.050$ & $-4.396 \pm 0.386$ & $0.23$ \\ 
 Spitzer $I3$ (Vega mag) & $5.80$ & $10.992 \pm 0.044$ & $-7.485 \pm 0.050$ & $-4.402 \pm 0.386$ & $0.23$ \\ 
 JWST MIRI F770W (Vega mag) & $7.70$ & $10.830 \pm 0.045$ & $-7.647 \pm 0.051$ & $-4.541 \pm 0.394$ & $0.23$ \\ 
 Spitzer $I4$ (Vega mag) & $8.00$ & $10.827 \pm 0.045$ & $-7.650 \pm 0.051$ & $-4.548 \pm 0.394$ & $0.23$ \\ 
 JWST MIRI F1000W (Vega mag) & $10.00$ & $10.786 \pm 0.046$ & $-7.691 \pm 0.052$ & $-4.638 \pm 0.401$ & $0.24$ \\ 
 JWST MIRI F1130W (Vega mag) & $11.30$ & $10.785 \pm 0.047$ & $-7.692 \pm 0.052$ & $-4.675 \pm 0.405$ & $0.24$ \\ 
 WISE $W3$ (Vega mag) & $12.00$ & $10.799 \pm 0.046$ & $-7.678 \pm 0.052$ & $-4.665 \pm 0.404$ & $0.24$ \\ 
 JWST MIRI F1280W (Vega mag) & $12.80$ & $10.719 \pm 0.047$ & $-7.758 \pm 0.053$ & $-4.702 \pm 0.407$ & $0.24$ \\ 
 JWST MIRI F1500W (Vega mag) & $15.00$ & $10.687 \pm 0.047$ & $-7.790 \pm 0.053$ & $-4.734 \pm 0.410$ & $0.24$ \\ 
 JWST MIRI F1800W (Vega mag) & $18.00$ & $10.669 \pm 0.047$ & $-7.808 \pm 0.053$ & $-4.760 \pm 0.412$ & $0.24$ \\ 
 JWST MIRI F2100W (Vega mag) & $21.00$ & $10.619 \pm 0.047$ & $-7.858 \pm 0.053$ & $-4.768 \pm 0.413$ & $0.24$ \\ 
 WISE $W4$ (Vega mag) & $22.00$ & $10.635 \pm 0.047$ & $-7.842 \pm 0.053$ & $-4.766 \pm 0.412$ & $0.24$ \\ 
 Spitzer $M1$ (Vega mag) & $24.00$ & $10.583 \pm 0.047$ & $-7.894 \pm 0.053$ & $-4.762 \pm 0.412$ & $0.24$ \\ 
 JWST MIRI F2550W (Vega mag) & $25.50$ & $10.586 \pm 0.047$ & $-7.891 \pm 0.053$ & $-4.751 \pm 0.410$ & $0.24$ \\ 
 \hline \hline
\end{tabular}
\tablecomments{The PLR slope $a_1$ from the linear fit in a form $m_\lambda = a_0 + a_1 \times (\log P-2.3)$. In the table, we provide the observational zero-point $a_{0,\mathrm{obs}}$ and absolute zero-point $a_{0,\mathrm{abs}}$ calculated with the distance modulus $\mu = 18.477$ mag \citep{2019Natur.567..200P}. The uncertainties of $a_{0,\mathrm{abs}}$ are a square root of a sum in quadrature of the $a_{0,\mathrm{obs}}$ uncertainty, the statistical ($0.09$ kpc), and systematical ($0.54$ kpc) LMC distance uncertainties.}
\end{table*}

\begin{table*}[ht!]
\caption{Parameters of synthetic PLRs for the C-rich Miras\label{tab:PLRparC}}
\vspace{0.2cm}
\footnotesize
\begin{tabular}{l|r|ccc|c}
\hline
\multicolumn{5}{c}{C-rich Miras} \\
\hline \hline
Survey and filter names & $\lambda_{\mathrm{eff}}$ ($\mu$m) & $a_{\mathrm{0, obs}}$ & $a_{\mathrm{0, abs}}$ & $a_1$ & RMS (mag) \\ 
\hline \hline
LSST $u$ (AB mag) & $0.37$ & $21.607 \pm 0.785$ & $3.130 \pm 0.786$ & $6.816 \pm 2.691$ & $2.98$ \\  LSST $g$ (AB mag) & $0.48$ & $17.910 \pm 0.613$ & $-0.567 \pm 0.613$ & $6.936 \pm 2.090$ & $2.33$ \\ 
 Gaia $G_{\mathrm{bp}}$ (Vega mag) & $0.51$ & $16.798 \pm 0.430$ & $-1.679 \pm 0.431$ & $6.269 \pm 1.463$ & $1.63$ \\ 
 OGLE $V$ (Vega mag) & $0.55$ & $17.141 \pm 0.520$ & $-1.336 \pm 0.521$ & $7.116 \pm 1.776$ & $1.98$ \\ 
 LSST $r$ (AB mag) & $0.63$ & $16.197 \pm 0.396$ & $-2.280 \pm 0.396$ & $6.084 \pm 1.346$ & $1.49$ \\ 
 Gaia $G$ (Vega mag) & $0.63$ & $14.893 \pm 0.310$ & $-3.584 \pm 0.311$ & $5.411 \pm 1.054$ & $1.17$ \\ 
 LSST $i$ (AB mag) & $0.76$ & $15.040 \pm 0.324$ & $-3.437 \pm 0.325$ & $5.593 \pm 1.103$ & $1.22$ \\ 
 Gaia $G_{\mathrm{rp}}$ (Vega mag) & $0.78$ & $14.231 \pm 0.295$ & $-4.246 \pm 0.296$ & $5.289 \pm 1.002$ & $1.11$ \\ 
 OGLE $I$ (Vega mag) & $0.80$ & $14.305 \pm 0.307$ & $-4.172 \pm 0.308$ & $5.436 \pm 1.043$ & $1.16$ \\ 
 LSST $z$ (AB mag) & $0.87$ & $14.343 \pm 0.288$ & $-4.134 \pm 0.289$ & $5.234 \pm 0.978$ & $1.09$ \\ 
 LSST $y$ (AB mag) & $0.97$ & $13.979 \pm 0.274$ & $-4.498 \pm 0.275$ & $4.724 \pm 0.934$ & $1.04$ \\ 
 VMC $Y$ (Vega mag) & $1.02$ & $13.259 \pm 0.246$ & $-5.218 \pm 0.247$ & $3.839 \pm 0.839$ & $0.93$ \\ 
 HST WFC3\_IR F110W (Vega mag) & $1.15$ & $12.645 \pm 0.241$ & $-5.832 \pm 0.242$ & $3.822 \pm 0.822$ & $0.92$ \\ 
 VMC $J$ (Vega mag) & $1.25$ & $12.490 \pm 0.238$ & $-5.987 \pm 0.239$ & $3.286 \pm 0.812$ & $0.90$ \\ 
 WFIRST WFI J129 (Vega mag) & $1.29$ & $12.317 \pm 0.239$ & $-6.160 \pm 0.241$ & $3.324 \pm 0.817$ & $0.91$ \\ 
 HST WFC3\_IR F140W (Vega mag) & $1.40$ & $12.089 \pm 0.229$ & $-6.388 \pm 0.230$ & $2.793 \pm 0.781$ & $0.87$ \\ 
 HST WFC3\_IR F160W (Vega mag) & $1.53$ & $11.866 \pm 0.192$ & $-6.611 \pm 0.193$ & $1.791 \pm 0.657$ & $0.73$ \\ 
 WFIRST WFI H158 (Vega mag) & $1.58$ & $11.826 \pm 0.188$ & $-6.651 \pm 0.189$ & $1.548 \pm 0.642$ & $0.71$ \\ 
 WFIRST WFI F184 (Vega mag) & $1.84$ & $11.590 \pm 0.147$ & $-6.887 \pm 0.149$ & $-0.027 \pm 0.507$ & $0.56$ \\ 
 JWST NIRCam F200W (Vega mag) & $1.99$ & $11.489 \pm 0.140$ & $-6.988 \pm 0.142$ & $-0.719 \pm 0.482$ & $0.53$ \\ 
 VMC $K_\mathrm{s}$ (Vega mag) & $2.20$ & $11.420 \pm 0.129$ & $-7.057 \pm 0.131$ & $-1.522 \pm 0.446$ & $0.49$ \\ 
 JWST NIRCam F277W (Vega mag) & $2.76$ & $10.822 \pm 0.103$ & $-7.655 \pm 0.105$ & $-3.811 \pm 0.356$ & $0.39$ \\ 
 WISE $W1$ (Vega mag) & $3.40$ & $11.134 \pm 0.091$ & $-7.343 \pm 0.094$ & $-5.093 \pm 0.315$ & $0.34$ \\ 
 JWST NIRCam F356W (Vega mag) & $3.57$ & $11.060 \pm 0.094$ & $-7.417 \pm 0.097$ & $-5.425 \pm 0.326$ & $0.35$ \\ 
 Spitzer $I1$ (Vega mag) & $3.60$ & $11.086 \pm 0.094$ & $-7.391 \pm 0.097$ & $-5.403 \pm 0.325$ & $0.35$ \\ 
 JWST NIRCam F444W (Vega mag) & $4.41$ & $10.957 \pm 0.122$ & $-7.520 \pm 0.124$ & $-6.689 \pm 0.419$ & $0.46$ \\ 
 Spitzer $I2$ (Vega mag) & $4.50$ & $10.987 \pm 0.123$ & $-7.490 \pm 0.125$ & $-6.799 \pm 0.423$ & $0.46$ \\ 
 WISE $W2$ (Vega mag) & $4.60$ & $10.989 \pm 0.124$ & $-7.488 \pm 0.127$ & $-6.938 \pm 0.428$ & $0.47$ \\ 
 JWST MIRI F560W (Vega mag) & $5.60$ & $10.891 \pm 0.143$ & $-7.586 \pm 0.145$ & $-7.686 \pm 0.492$ & $0.54$ \\ 
 Spitzer $I3$ (Vega mag) & $5.80$ & $10.904 \pm 0.144$ & $-7.573 \pm 0.146$ & $-7.728 \pm 0.495$ & $0.54$ \\ 
 JWST MIRI F770W (Vega mag) & $7.70$ & $10.860 \pm 0.181$ & $-7.617 \pm 0.182$ & $-8.722 \pm 0.626$ & $0.69$ \\ 
 Spitzer $I4$ (Vega mag) & $8.00$ & $10.860 \pm 0.182$ & $-7.617 \pm 0.183$ & $-8.771 \pm 0.629$ & $0.69$ \\ 
 JWST MIRI F1000W (Vega mag) & $10.00$ & $10.913 \pm 0.185$ & $-7.564 \pm 0.186$ & $-9.470 \pm 0.638$ & $0.70$ \\ 
 JWST MIRI F1130W (Vega mag) & $11.30$ & $10.926 \pm 0.190$ & $-7.551 \pm 0.191$ & $-9.707 \pm 0.655$ & $0.72$ \\ 
 WISE $W3$ (Vega mag) & $12.00$ & $10.880 \pm 0.196$ & $-7.597 \pm 0.198$ & $-9.529 \pm 0.680$ & $0.74$ \\ 
 JWST MIRI F1280W (Vega mag) & $12.80$ & $10.810 \pm 0.201$ & $-7.667 \pm 0.202$ & $-9.741 \pm 0.695$ & $0.76$ \\ 
 JWST MIRI F1500W (Vega mag) & $15.00$ & $10.789 \pm 0.206$ & $-7.688 \pm 0.207$ & $-9.944 \pm 0.712$ & $0.78$ \\ 
 JWST MIRI F1800W (Vega mag) & $18.00$ & $10.781 \pm 0.210$ & $-7.696 \pm 0.212$ & $-10.129 \pm 0.728$ & $0.80$ \\ 
 JWST MIRI F2100W (Vega mag) & $21.00$ & $10.739 \pm 0.213$ & $-7.738 \pm 0.214$ & $-10.241 \pm 0.738$ & $0.81$ \\ 
 WISE $W4$ (Vega mag) & $22.00$ & $10.759 \pm 0.214$ & $-7.718 \pm 0.216$ & $-10.288 \pm 0.743$ & $0.81$ \\ 
 Spitzer $M1$ (Vega mag) & $24.00$ & $10.712 \pm 0.215$ & $-7.765 \pm 0.217$ & $-10.323 \pm 0.746$ & $0.82$ \\ 
 JWST MIRI F2550W (Vega mag) & $25.50$ & $10.721 \pm 0.217$ & $-7.756 \pm 0.218$ & $-10.365 \pm 0.751$ & $0.82$ \\ 
 \hline \hline
\end{tabular}
\tablecomments{The PLR slope $a_1$ from the linear fit in a form $m_\lambda = a_0 + a_1 \times (\log P-2.3)$. In the table, we provide the observational zero-point $a_{0,\mathrm{obs}}$ and absolute zero-point $a_{0,\mathrm{abs}}$ calculated with the distance modulus $\mu = 18.477$ mag \citep{2019Natur.567..200P}. The uncertainties of $a_{0,\mathrm{abs}}$ are a square root of a sum in quadrature of the $a_{0,\mathrm{obs}}$ uncertainty, the statistical ($0.09$ kpc), and systematical ($0.54$ kpc) LMC distance uncertainties.}
\end{table*}

\begin{figure*}
\centering
\includegraphics[scale=0.5]{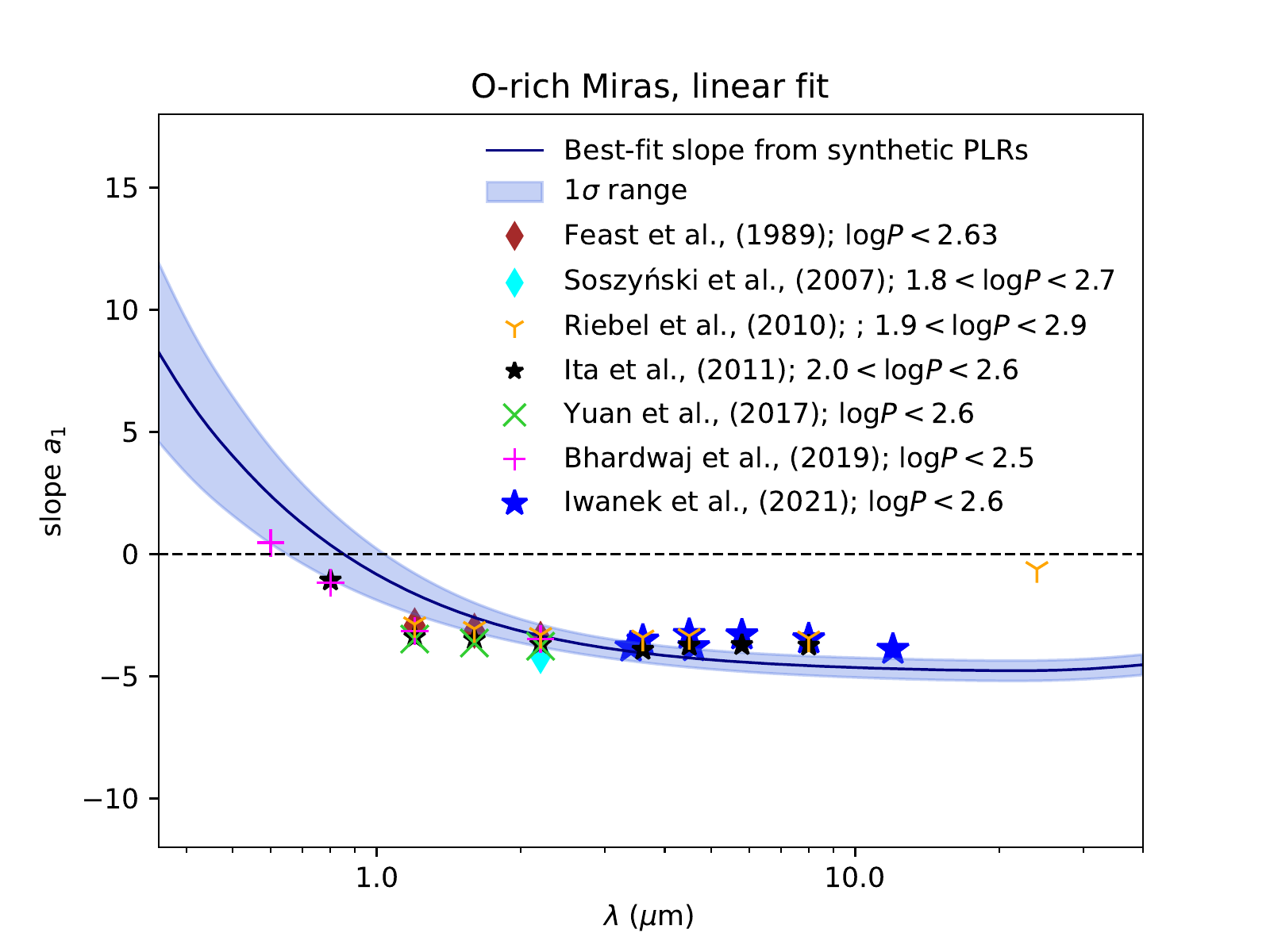}
\includegraphics[scale=0.5]{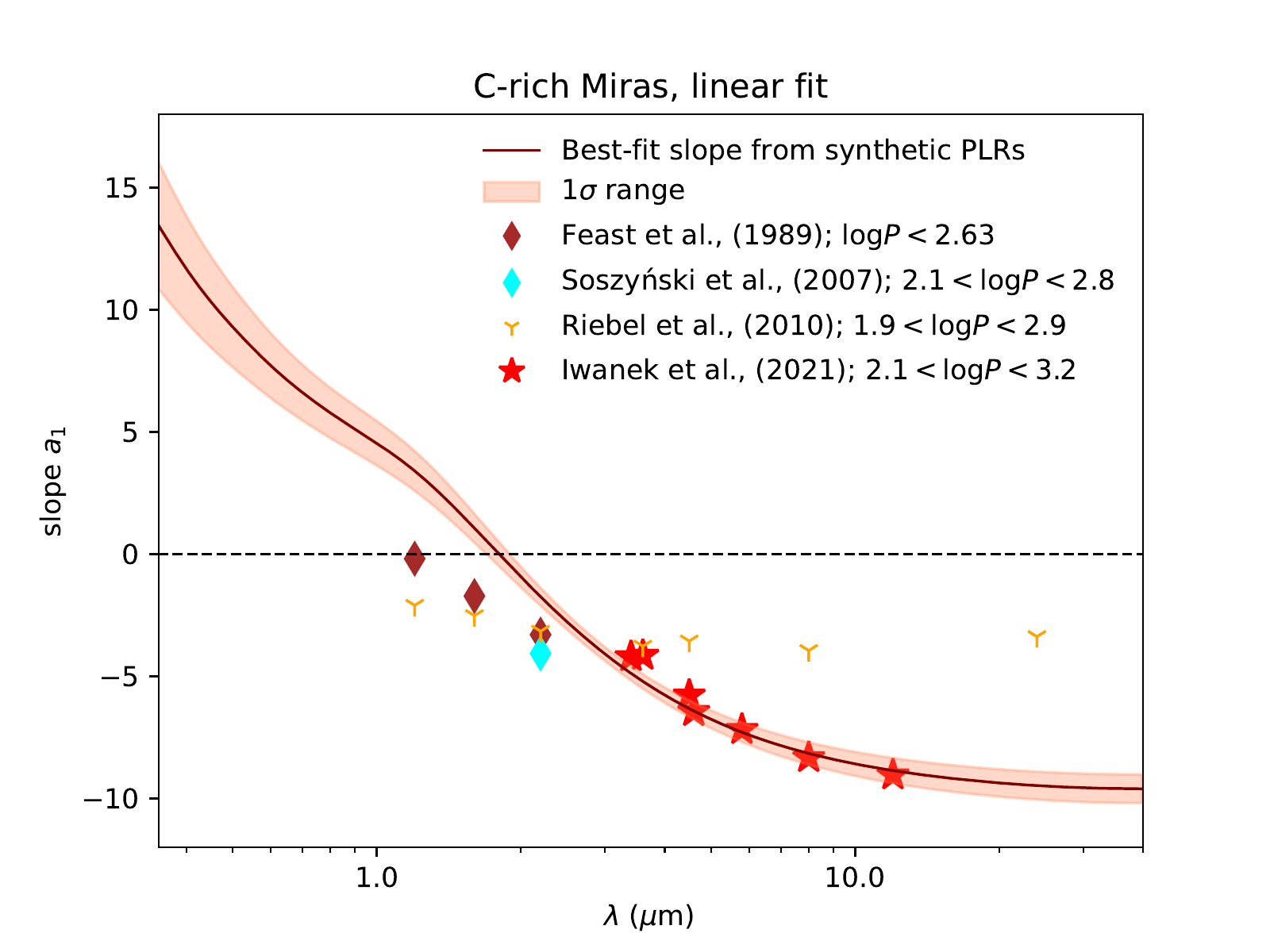}
\caption{The synthetic PLR slope ($a_1$) from a linear fit ($m = a_0+a_1 \times (\log P-2.3)$) as a function of wavelength is shown as the blue (red) band in the left (right) panel for the O-rich (C-rich) Miras from the golden sample. With symbols we mark measurements from publications provided in the legend.}
\label{fig:slope-lambda}
\end{figure*}

From the synthetic SEDs and without using particular filters, we also calculated the PLR slope as a function of wavelength in a range of 0.1--40 microns (Table~\ref{tab:synthetic_slope}). Indeed, the PLR slope changes smoothly from the positive slope in the UV to the negative slope in the near- and mid-IR wavelengths. In Figure~\ref{fig:slope-lambda}, we present the relation between the synthetic slope and wavelength for the O-rich Miras (left panel) and the C-rich Miras (right panel). The shaded band in both panels corresponds to the uncertainty in the slope measurement from the synthetic magnitudes. In both panels, we also presented measurements of the PLR slope from the literature, that include \cite{1989MNRAS.241..375F}, \cite{2007AcA....57..201S}, \cite{2010ApJ...723.1195R}, \cite{2011MNRAS.412.2345I}, \cite{2017AJ....153..170Y}, \cite{2019ApJ...884...20B}, and \cite{2020arXiv201212910I}. 

For the O-rich Miras (left panel of Figure~\ref{fig:slope-lambda}), the vast majority of measurements reported in the literature follow our synthetic PLR slope-wavelength relation. 
Note, however, that our synthetic relation serves here as a guidance only, because our data is limited to a small number of \goldnum stars, spanning a period range of $\log P\approx2.0$ to 2.8, and well-described by a linear relation. On the other hand, a large sample of the O-rich Miras shows a clear PLR relation that is not linear but rather parabolic (\citealt{2020arXiv201212910I}).

\begin{table*}[t!]
\scriptsize
\begin{center}
\caption{Basic parameters of the golden sample Miras.}
\label{tab:goldsample}
\vspace{0.2cm}
\begin{tabular}{lcccccccccr}
\hline \hline 
Number & RA $(^h:^m:^s)$ & Decl. $(^\circ:^m:^s)$& type & $P$ (d) & $T^{\mathrm{star}}$ (K) & $T^{\mathrm{dust}}$ (K) & LSST $u$ (mag)  & LSST $g$ (mag) & \ldots & JWST F2550W (mag) \\ \hline \hline 
08058 & 04:56:54.06 & $-$67:34:11.8 & C & 383.59 & 2238 & 1095 & 23.081 & 19.375 &  \ldots &  6.849 \\
08424 & 04:57:21.71 & $-$67:21:28.0 & C & 514.71 & 2055 & 909 & 25.130 & 21.005 &  \ldots & 5.445 \\
09268 & 04:58:25.31 & $-$68:08:36.3 & C &  336.20 &  2906 &  1340 &  19.432 &  16.796 &  \ldots &  9.275 \\
10812 & 05:00:11.95 & $-$67:40:10.7 & C &  427.05 &  1865 &  963 & 26.779 &  22.127 &  \ldots &  6.698 \\
12829 & 05:02:11.27 & $-$68:12:15.5 & C & 231.10 &  2630 & 0 & 20.151 & 17.142 & \ldots & 9.583 \\
13563 & 05:02:52.86 & $-$67:07:39.1 & C & 317.46 & 2695 & 1522 & 20.874 & 17.957&  \ldots & 8.636 \\
13815 & 05:03:06.06 & $-$67:58:00.5 & C & 442.79 &  1837 & 1011 & 34.147 & 26.729&  \ldots &  7.603 \\
14860 & 05:03:59.29 & $-$68:11:35.9 & C & 201.87 & 2796 & 1514 & 23.164 & 19.313 &  \ldots &  9.515 \\
15353 & 05:04:26.15 & $-$68:18:43.2 & C & 305.74 & 2550 & 988 & 22.995 & 19.213&  \ldots & 9.595 \\
16684 & 05:05:28.00 & $-$68:09:19.6 & C & 462.31 & 1835  & 1214 &  23.535 & 20.327 & \ldots & 12.187 \\
\vdots & \vdots & \vdots &  \vdots & \vdots & \vdots & \vdots & \vdots & \vdots & $\ddots$ &  \vdots  \\
82526 & 05:44:13.40 & $-$69:03:20.4 & O & 140.28 & 2479 & 0 & 22.729 & 19.516&  \ldots & 11.361 \\
\hline \hline
\end{tabular}
\end{center}
\tablecomments{We provide the coordinates, surface chemistry classification, pulsation periods, temperatures for hotter and cooler components from the SEDs fitting, and 42 synthetic, calibrated (extinction-free) magnitudes in the analyzed bands from existing and future sky surveys. The full star ID is as in the original catalog \citep{2009AcA....59..239S} and is composed of OGLE-LMC-LPV-, followed by the number. Table rows are sorted by the star ID, while columns with magnitudes are sorted by wavelength. The $T^{\mathrm{dust}}$ for O-rich Miras are reported as 0 in the table, as SEDs for these stars required only hotter components (for more details see Section \ref{sec:SED}). This table is available in its entirety in a machine-readable form in the online journal. A portion is shown here for guidance regarding its form and content.}

\end{table*}

Because SEDs of the C-rich Miras are typically well-described by two Planck components, the hot one and the cooler one (as in Figure~\ref{fig:SED_C}), this combination may somewhat impact the PLR slope-wavelength relation. In particular, the cooler component that is sparsely present in the O-rich Miras in the near-IR and mid-IR, for the C-rich Miras it is significant at these wavelengths (so adds typically a significant fraction of light in these bands). From the right panel of Figure~\ref{fig:slope-lambda}, we can see that measurements reported in \cite{1989MNRAS.241..375F}, \cite{2007AcA....57..201S}, and \cite{2010ApJ...723.1195R} show a clear departure from the synthetic PLR slope-wavelength relation. On the other hand, measurements from \citet{2020arXiv201212910I} seem to follow that relation closely.

The list of \goldnum Miras from the golden sample, that include their coordinates, surface chemistry classification, pulsation periods, temperatures and 42 synthetic, calibrated (extinction-free) magnitudes from the existing and future sky surveys are presented in Table \ref{tab:goldsample}.

\newpage
\section{Miras on CMDs and CCDs}
\label{sec:CMDs}

As described in Section~\ref{sec:data}, our full sample of Miras consists of 1663 stars with light curves in the OGLE survey, 595 counterparts in the VMC data set, and 1311 identifications in the WISE survey. 
We analysed the location of these stars in the OGLE, VMC and WISE color-magnitude diagrams (CMD) and color-color diagrams (CCD).

\begin{figure*}
\centering
\includegraphics[scale=0.5]{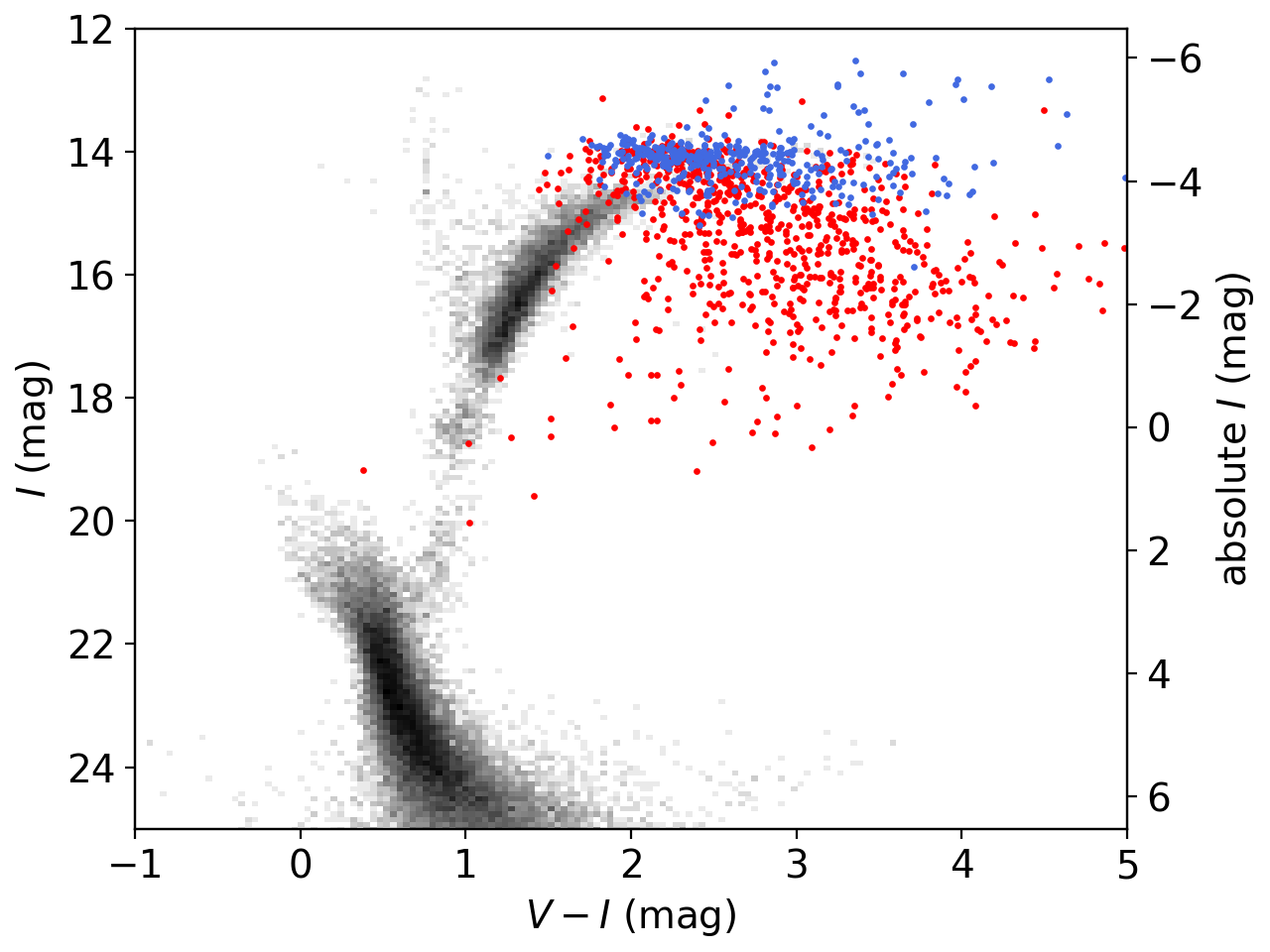} 
\includegraphics[scale=0.5]{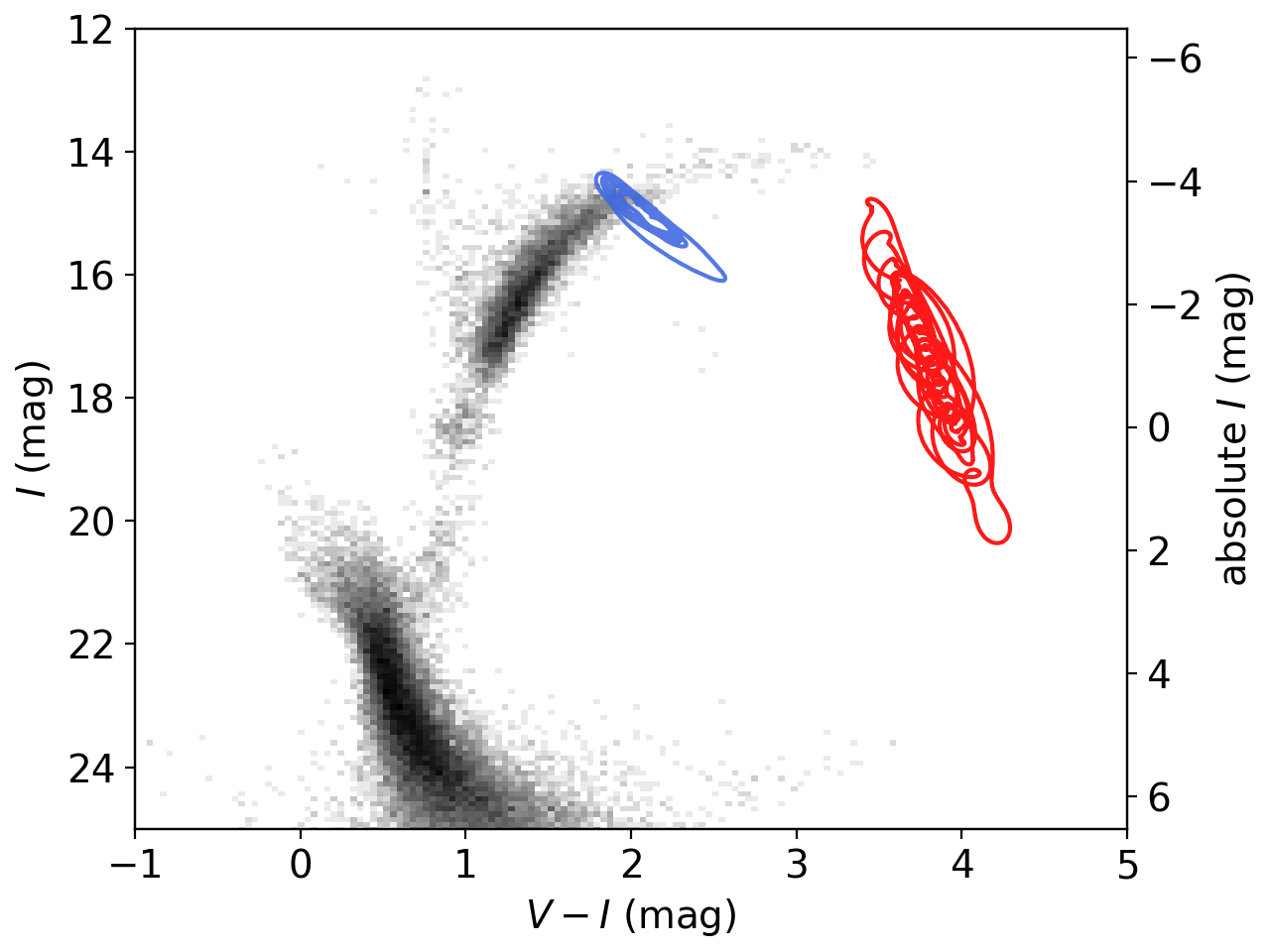}\\
\caption{Extinction-corrected color-magnitude diagram in the OGLE $VI$ filters is shown (grey 2D histogram), where $I<18$~mag data are from OGLE and $I>18$~mag data are from HST. The left panel shows 1663 LMC Miras divided into O-rich (blue) and C-rich (red). The right panel shows the motion of a bolometrically faint, 5200~$L_\odot$ O-rich Mira (blue, Mira OGLE ID: OGLE-LMC-LPV-36521) and a bolometrically luminous, 46000~$L_\odot$ C-rich Mira (red, Mira OGLE ID: OGLE-LMC-LPV-08424) on the CMD. Loops for the O-rich Mira span $11$ pulsation periods, while loops for the C-rich Mira span $13$ pulsation periods.}
\label{fig:CMD_OGLE}
\end{figure*}

In both panels of Figure~\ref{fig:CMD_OGLE} as the grey 2D histograms, we present field stars where for $I<18$~mag we used the OGLE dataset and for the $I>18$~mag we used the HST $F555W$ and $F814W$ data. Because of that we can clearly see a density change above the red clump stars with $(V-I, I)\approx (1.0,18.48)$~mag. In the left panel of Figure~\ref{fig:CMD_OGLE}, with blue (red) dots we present the O-rich (C-rich) Miras. It is clear that on average in the $I$-band the O-rich Miras appear to be brighter than the C-rich Miras. In the right panel of Figure~\ref{fig:CMD_OGLE}, we show the motion on the CMD (loops) of two Miras during multiple pulsation cycles: a bolometrically faint, 5200~$L_\odot$ O-rich Mira (blue, Mira OGLE ID: OGLE-LMC-LPV-36521) and a bolometrically luminous, 46000~$L_\odot$ C-rich Mira (red, Mira OGLE ID: OGLE-LMC-LPV-08424). The red loops after each period seem to be somewhat shifted along the long axis as this star significantly changes the mean brightness with time.

\begin{figure*}
\centering
\includegraphics[scale=0.5]{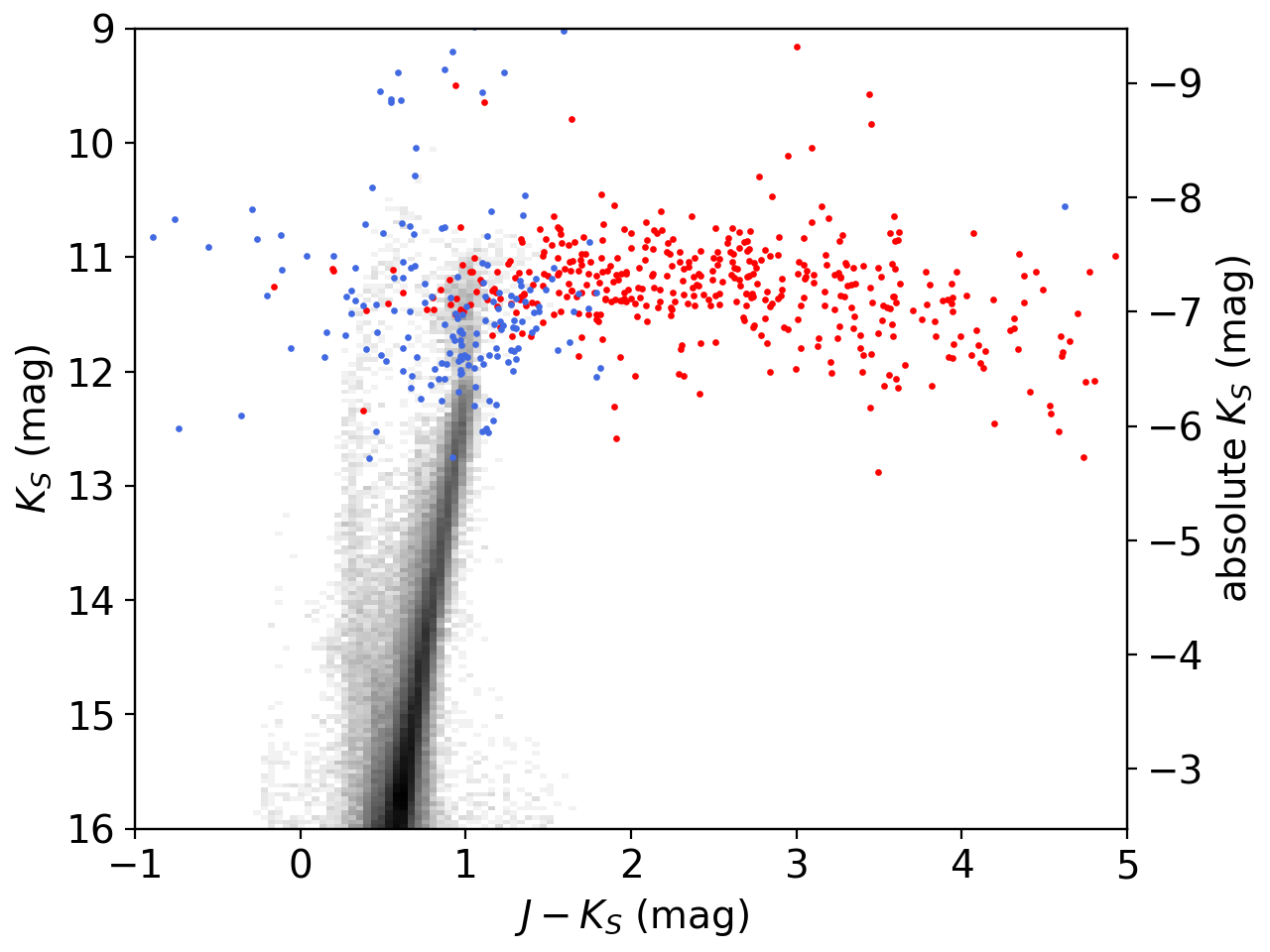} 
\includegraphics[scale=0.5]{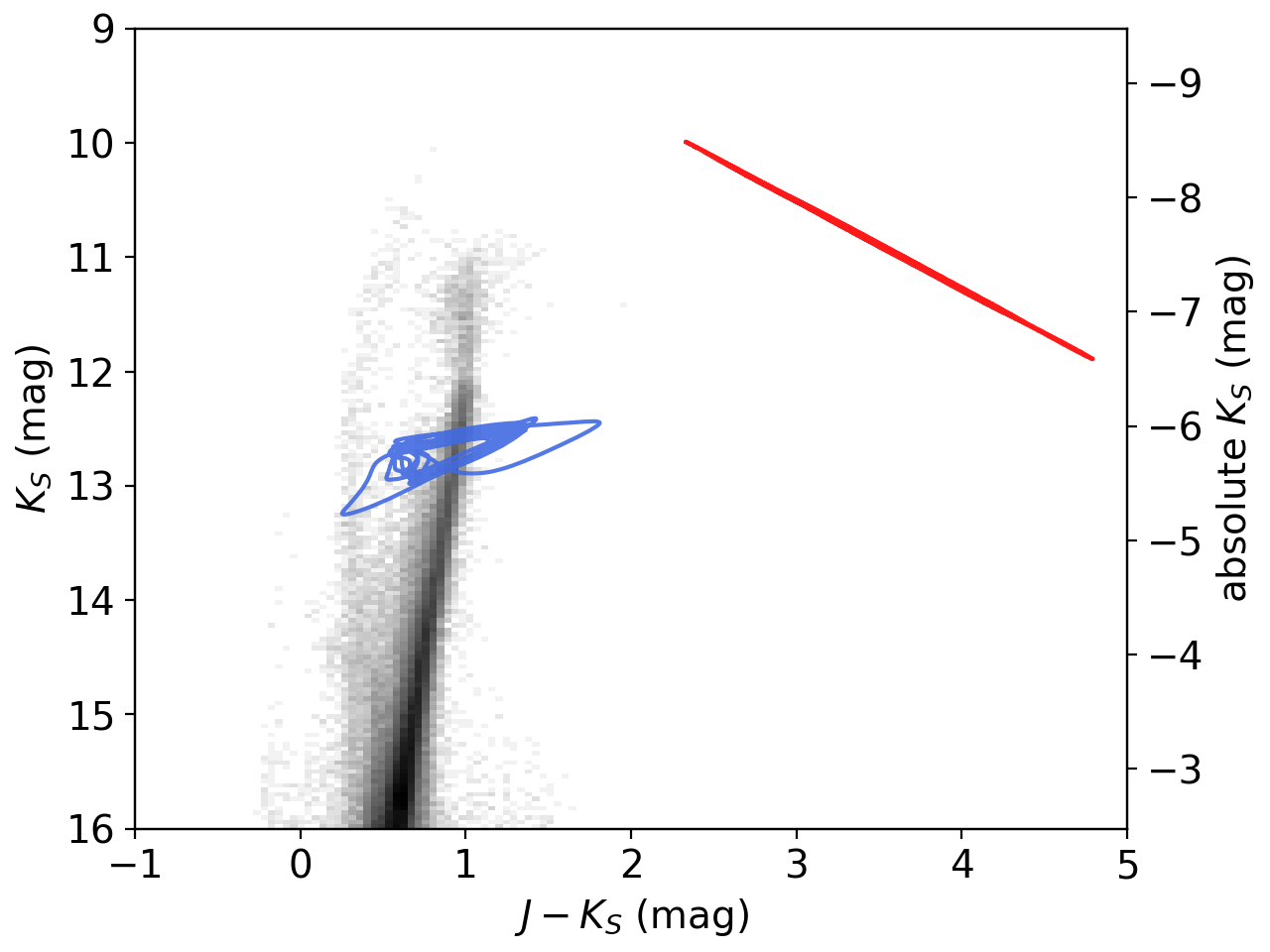}\\
\includegraphics[scale=0.5]{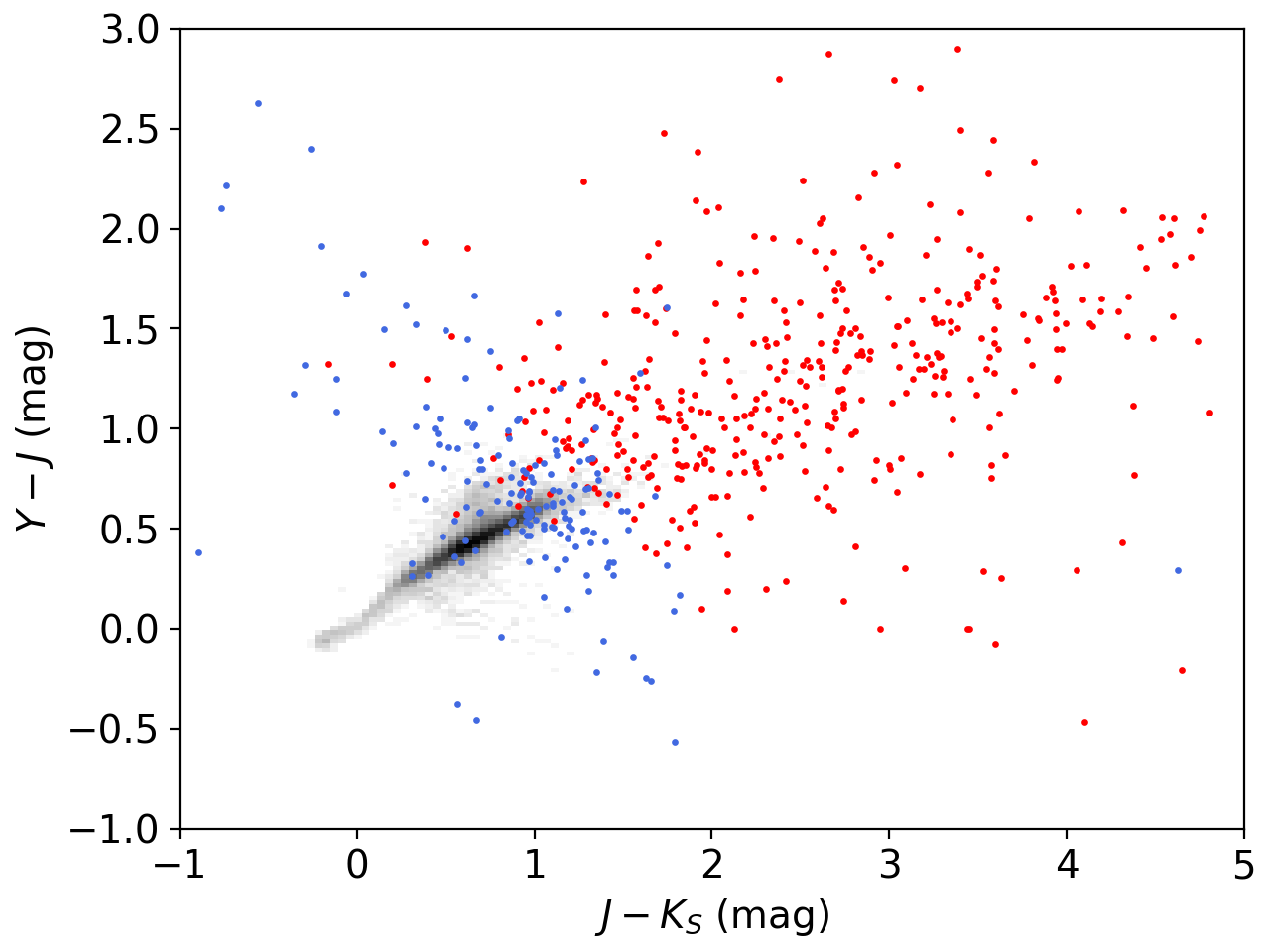} 
\includegraphics[scale=0.5]{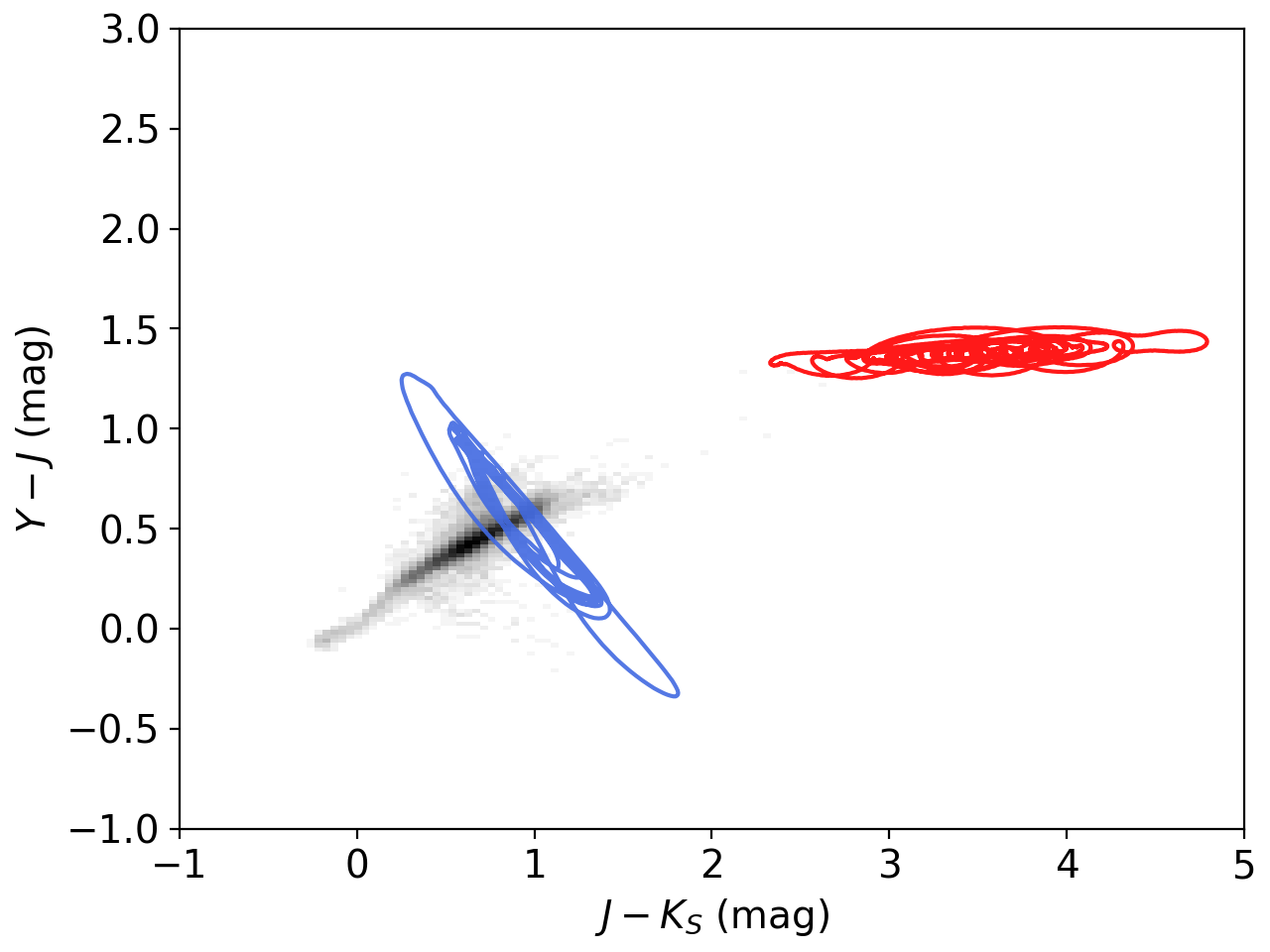}
\caption{Extinction-corrected color-magnitude diagram in the VMC $JK_\mathrm{s}$ filters is shown in the top row (grey 2D histogram) and the color-color diagram in the $YJK_\mathrm{s}$ filters in the bottom row. The left column shows 595 LMC Miras divided into O-rich (blue) and C-rich (red). The right column shows the motion of a bolometrically faint, 5200~$L_\odot$ O-rich Mira (blue, Mira OGLE ID: OGLE-LMC-LPV-36521) and a bolometrically luminous, 46000~$L_\odot$ C-rich Mira (red, Mira OGLE ID: OGLE-LMC-LPV-08424) on the CMD. Loops for the O-rich Mira span $11$ pulsation periods, while loops for the C-rich Mira span $13$ pulsation periods.}
\label{fig:CMD_VMC}
\end{figure*}

In Figure~\ref{fig:CMD_VMC}, we present the location of Miras in the CMDs (top row) and CCDs (bottom row) for the VMC survey with $Y$, $J$ and $K_\mathrm{s}$ filters. The field stars are, again, presented as 2D grey histograms, while the 595 matched Miras are presented in the left column and the motion of the two Miras are presented in the right column. Both the O- and C-rich Miras are very luminous in the $K_\mathrm{s}$-band, with the absolute magnitude $K_\mathrm{s}\approx-7$~mag. Because the O-rich Miras are less affected by dust, they lie on the top of the locus of the field stars. The C-rich Miras, on the other hand, are affected by the dust (that adds IR light) and their location is shifted towards redder $J-K_\mathrm{s}$ colors. In the right-top panel of Figure~\ref{fig:CMD_VMC}, we present the motion of the two fore-mentioned Miras: a bolometrically faint, 5200~$L_\odot$ O-rich Mira (blue, Mira OGLE ID: OGLE-LMC-LPV-36521) and a bolometrically luminous, 46000~$L_\odot$ C-rich Mira (red, Mira OGLE ID: OGLE-LMC-LPV-08424). The track of the red C-rich Mira appears as to be a straight line, because the phase-lag between the $J$- and $K_\mathrm{s}$-band light curves is tiny, 0.8 days or 0.0016 in phase. 

\begin{figure*}
\centering
\includegraphics[scale=0.5]{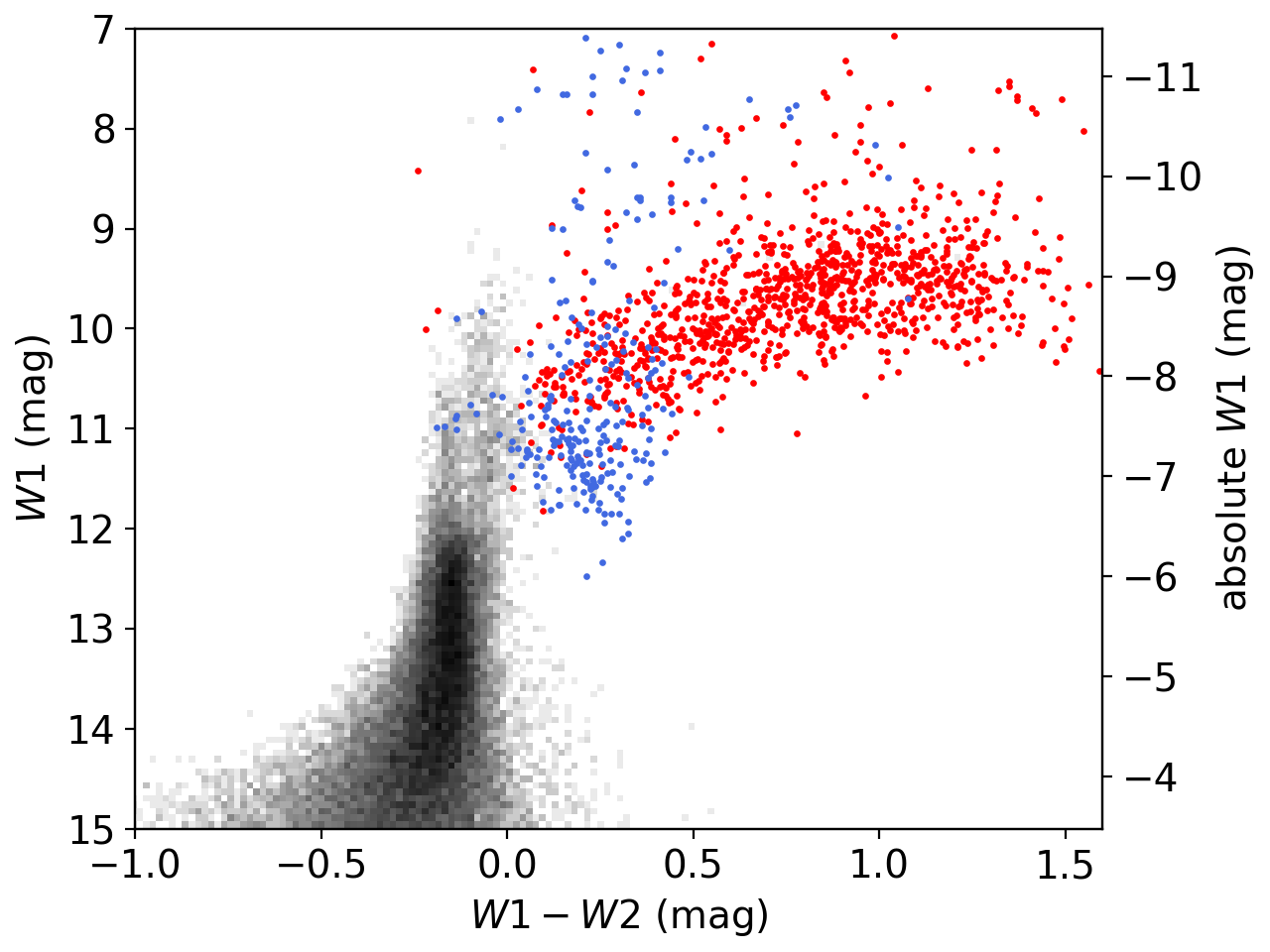} 
\includegraphics[scale=0.5]{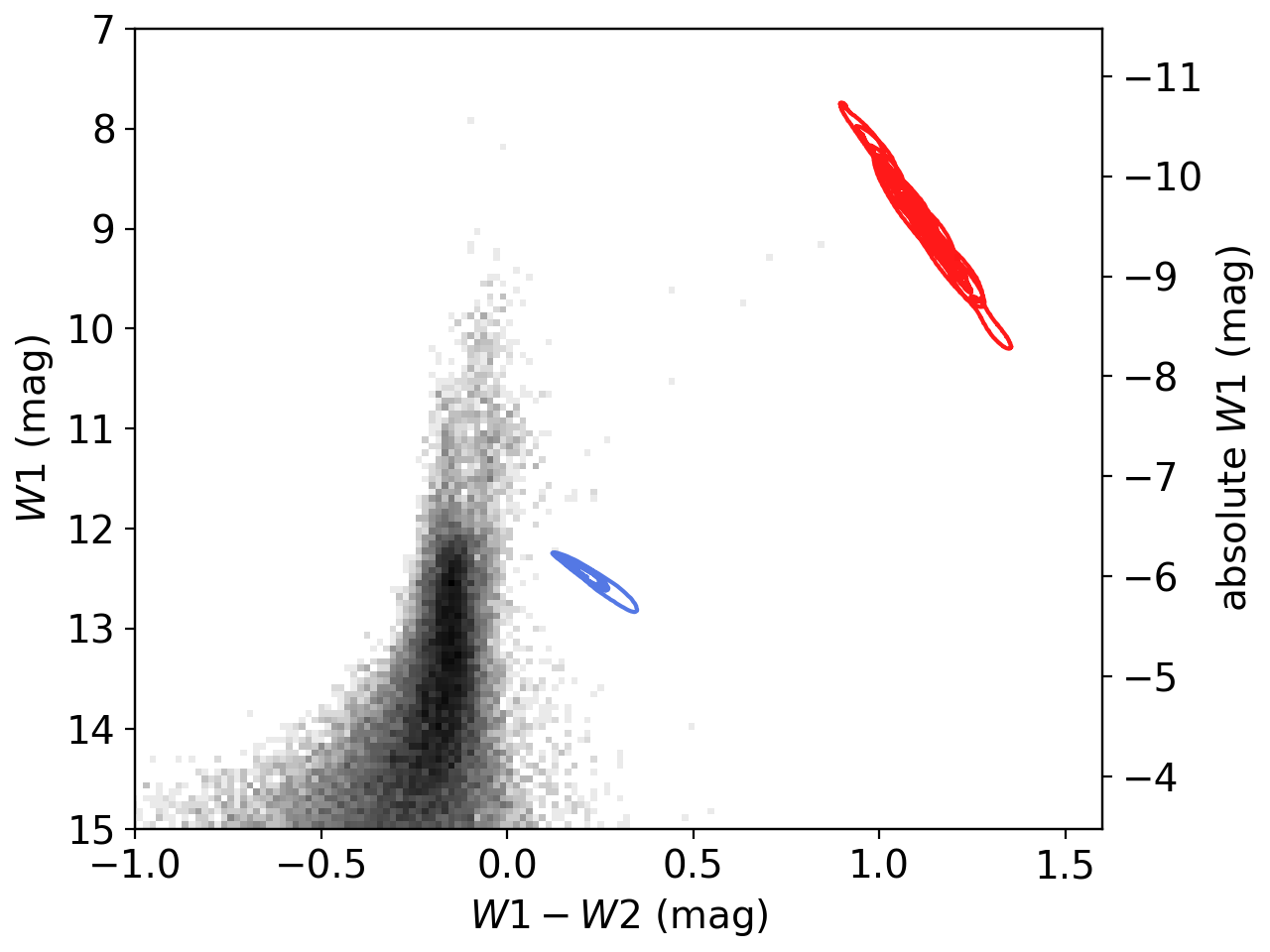}\\
\includegraphics[scale=0.5]{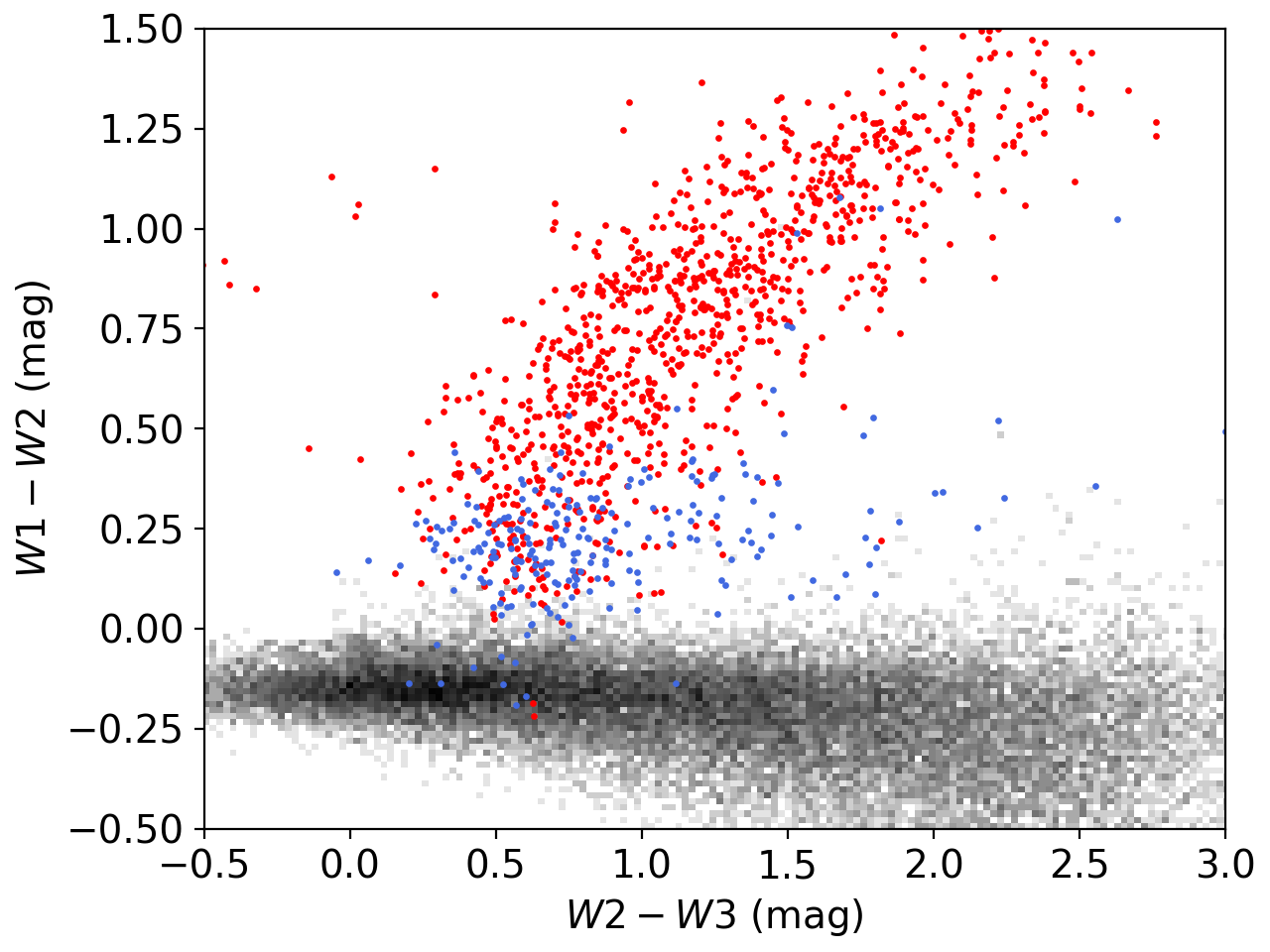}
\caption{Extinction-corrected color-magnitude diagram in the WISE $W1W2$ filters is shown in top panels (grey 2D histogram), and the color-color diagram in the $W1W2W3$ filters in the bottom panel. Top-left and bottom panels shows 1311 LMC Miras divided into O-rich (blue) and C-rich (red). Top-right panel shows the  motion of a bolometrically faint, 5200~$L_\odot$ O-rich Mira (blue, Mira OGLE ID: OGLE-LMC-LPV-36521) and a bolometrically luminous, 46000~$L_\odot$ C-rich Mira (red, Mira OGLE ID: OGLE-LMC-LPV-08424) on the CMD.}
\label{fig:CMD_WISE}
\end{figure*}

In Figure~\ref{fig:CMD_WISE}, we present the location of Miras in the CMDs (top panel) and motions on CMD of the two fore-mentioned Miras: a bolometrically faint, 5200~$L_\odot$ O-rich Mira (blue, Mira OGLE ID: OGLE-LMC-LPV-36521) and a bolometrically luminous, 46000~$L_\odot$ C-rich Mira (red, Mira OGLE ID: OGLE-LMC-LPV-08424; middle panel), and CCDs (bottom panel) for the WISE survey. In the top panel, the red C-rich Miras extend significantly from the stellar locus toward red $W1-W2$ colors, while the O-rich Miras appear to be on average fainter and bluer than the C-rich Miras. In the bottom panel of Figure~\ref{fig:CMD_WISE}, the red points (C-rich Miras) are scattered along the diagonal, what corresponds to the decreasing black body temperature towards the top-right corner. The coolest Miras with colors in the vicinity of $(W2-W3$, $W1-W2)\approx($2.5,1.2$)$ enter the color area occupied by active galactic nuclei (e.g., Figure~2 in \citealt{2014MNRAS.442.3361N}).


\section{Conclusions}
\label{sec:summary}

In this paper, we studied the variability of 1663  Mira-type stars in the LMC at a wide range of wavelengths (0.5--24 microns) and timescales (up to 25 years). We modeled the OGLE $I$-band data, with the median of 1310 epochs and up to 25 years span, using a Gaussian Process model that included the periodic component, slowly changing mean component, and a high-frequency stochastic component. We then fitted this high-quality model to light curves in 13 bands ($V$, $Y$, $J$, $K_\mathrm{s}$, Spitzer IRAC 3.6--8.0 micron, and WISE 3.5--22 micron data).

The fitting procedure provided us with the variability amplitude ratio $R$ (Figure~\ref{fig:var_amp}) and phase-lag $\phi$ (Figure~\ref{fig:var_phase} of other bands as compared to $I$-band. For both the O-rich and C-rich Miras the variability amplitude ratio decreases with the increasing wavelength. The phase-lag, on the other hand, seems to weakly increase with the increasing wavelength, where the $V$-band light curves seem to lead the $I$-band, with the near-IR and mid-IR light curves lagging.

We selected a subsample of \goldnum Mira stars (the golden sample) with high quality light curves in the $V$, $I$, $Y$, $J$, $K_\mathrm{s}$, $W1$, and $W2$ bands. For each star, we scaled the high-quality $I$-band model 
to these bands and created 6000 SEDs spanning 6000 days. Each SED was model with either a single or double Planck functions. With a large number of SED models as a function of time, we studied
the temperature and flux changes as a function of time. As a rule of thumb the O-rich Miras are well-described by a single Planck function (hence a lack of dust presence), while the C-rich Miras do require two Planck functions, what we interpret as the presence of both the star with higher temperature and dust component with a lower temperature. The SEDs of O-rich Miras peak at about 1 micron with a typical (i.e., median) temperature of 2900 K and the C-rich Miras show the SED peak closer to 2 microns with the stellar temperature of 2300 K and the dust temperature of 1000 K. 

The median bolometric luminosity of the golden sample Miras is 16000~$L_\odot$, with a range from 800~$L_\odot$ (for the C-rich Mira OGLE-LMC-LPV-38689 with $P=450.69$~d) to nearly 50400~$L_\odot$ (for the C-rich Mira OGLE-LMC-LPV-37168 with $P=555.60$~d).

From synthetic SED-based light curves, we derived the synthetic variability amplitude ratio $R$ (Table~\ref{tab:amp_ratio}) and phase-lag $\phi$ (Table~\ref{tab:PhaseLag}) as a function of wavelength (0.1--40 microns) for the O-rich and C-rich Miras in the golden sample, shown as color bands in Figures~\ref{fig:var_amp} and \ref{fig:var_phase}.

For the golden sample stars, we also calculated synthetic PLRs for 42 bands (Tables~\ref{tab:PLRparO} and \ref{tab:PLRparC}, Figures~\ref{fig:PLR_O1}, \ref{fig:PLR_O2}, \ref{fig:PLR_C1}, and \ref{fig:PLR_C2}) using filters from major current or future surveys (Figure~\ref{fig:filters}) and the mean SEDs for our golden sample. The calibrated absolute and observed magnitude zero points ($a_0$) for both O-rich and C-rich Miras at periods of 200 days ($\log P=2.3$) and PLR slopes ($a_1$) are provided in Tables~\ref{tab:PLRparO} and \ref{tab:PLRparC}, respectively. The 42 filters include The Vera C. Rubin Observatory ($u$, $g$, $r$, $i$, $z$, $y$), Gaia ($G$, $G_{bp}$, $G_{rp}$), OGLE ($V$, $I$), VMC ($Y$, $J$, $K_\mathrm{s}$), HST ($F110W$, $F140W$, $F160W$), The Nancy Grace Roman Space Telescope ($J129$, $H158$, $F184$), JWST ($F200W$, $F277W$, $F356W$, $F444W$, $F560$, $F770W$, $F1000$, $F1130W$, $F1280$, $F1500W$, $F1800W$, $F2100W$, $F2550W$), Spitzer (3.6, 4.5, 5.8, 8.0, 24 micron), and WISE (3.4, 4.6, 12, 22 micron). 

We also studied the change of PLR slope ($a_1$) as a function of wavelength from 0.1 to 40 microns (Table~\ref{tab:synthetic_slope}, Figure~\ref{fig:slope-lambda}). The PLR slope for both O-rich and C-rich Miras strongly decreases with the increasing wavelength, with roughly zero slope in optical and plateauing in mid-IR. We therefore confirm findings from preceding papers that PLR slopes in optical are generally flat (e.g., \citealt{2019ApJ...884...20B}) and that they are sloped in near-IR and mid-IR (e.g., \citealt{1989MNRAS.241..375F, 2007AcA....57..201S, 2010ApJ...723.1195R, 2011MNRAS.412.2345I, 2017AJ....153..170Y, 2019ApJ...884...20B, 2020arXiv201212910I}).

Finally, we present the locations and motions as a function of time, phase, or temperature for the O-rich and C-rich Miras on both color-magnitude diagrams and color-color diagrams for the OGLE, VMC and WISE surveys.

\section{Future} \label{sec:future}

Miras are very bright stars in the near- and mid-IR, with an approximate absolute magnitude of $-6$~mag/$-7$~mag at about 1--2 microns (at the SED peak). This means that Miras can be observed to at least multi-Mpc distances in the Universe.

We will now discuss the usability of Miras as a distance indicator in the JWST NIRCam F200W and F444W (2.0 and 4.4 micron) filters. 
As our model galaxy, we will use M33 with the distance modulus $24.57\pm0.05$~mag, or $820_{-19}^{+20}$~kpc (\citealt{2012ApJ...758...11C}), with about 1 deg in apparent size on the sky. 

As estimated from our analyses, the O-rich Miras have the absolute magnitude at $P=200$~days ($\log P=2.3$) of $-6.55\pm0.04$~mag in the F200W filter and  $-7.42\pm0.04$~mag in the F444W filter, while  the C-rich Miras have $-6.96\pm0.01$~mag and $-7.49\pm0.01$~mag, respectively (Tables~\ref{tab:PLRparO} and \ref{tab:PLRparC}). At the M33 distance they will have the observed brightness at $P=200$~days ($\log P=2.3$) of $18.02$~mag and $17.15$~mag for the O-rich Miras, and $17.61$~mag and $17.08$~mag for the C-rich Miras in the F200W and F444W filters, respectively.

We used the \cite{2017AJ....153..170Y} catalog of 1848 Miras in M33 to derive the nearest projected distance between them. The caveat in our rough estimation here is that this sample is approximately 80--90\% complete with a 12\% impurity. Also note, there will be many IR sources in a galaxy that are not Miras, what will certainly lead to severe blending in the images. Using the difference images analaysis method (\citealt{1998ApJ...503..325A,2000AcA....50..421W}), however, the problem of blending may be alleviated as
on the difference images all non-variable sources vanish and what is left are only variable and/or moving sources.

The point spread function (PSF) full width at half maximum (FWHM) in the F200W filter is $0.066$ arcsec or $2.14$ pixel and $0.145$ arcsec or $2.302$ pixel in F444W.
As a rough guide, we calculated separations between all pairs of Miras from \cite{2017AJ....153..170Y} and found a minimum value of $0.12$ arcsec or $0.05$ kpc. For the conservative estimation, we assume we can resolve stars separated by twice the FWHM (0.29 arcsec in F444W). We put our model M33 galaxy at a distance, where the typical separation between Miras is equal to our conservative resolution of 0.29 arcsec. This is about 42 times further away than its present distance (so approximately 34 Mpc), what would make these Miras fainter by 8.12~mag, meaning the observed magnitude of approximately 25 mag. This may be too faint of a magnitude for a reasonable JWST observing program.

Let us discuss this issue from the observed magnitude perspective. We assume that we would like to detect Miras with $>0.3$ mag variability amplitude at 2--4 microns ($>0.8$ mag in the $I$-band times the variability amplitude ratio of 0.4) at a distance of 10 Mpc (distance modulus 30~mag). To detect 0.3 mag variability amplitude reliably, we require the photometric uncertainties below 0.1 mag or the signal-to-ratio above 25. We used the JWST Exposure Time Calculator (\citealt{2016SPIE.9910E..16P}) to find that in a 12 minute integration we should reach the signal-to-noise ratio of 25 for a 23~mag Mira in F200W, and 4 minute exposure time in F444W to reach the signal-to-ratio of 30. 

As a rule of thumb a JWST observing program for a 10~Mpc galaxy would require about 30 observations spanning about 2 years ($\sim$2 pulsation periods) to reliably measure periods of Miras, their mean magnitudes, and colors to separate between the O-rich and C-rich Miras. This would mean a total of 6-to-7-hour JWST observing program in both F200W and F444W filters.

The mock M33 galaxy at a distance of 10~Mpc means a 12 times greater distance than the real one, what also means 12 smaller apparent size on the sky. Our mock galaxy would have a size of about 5 arcmin on the sky. The field-of-view of NIRCam is $2.1\times2.1$ arcmin, significantly less than the size of this mock galaxy. Given the high expected interest in the JWST time, however, a monitoring program of a single pointing per such a galaxy will have to suffice.

\begin{acknowledgments}
We are grateful to the anonymous referee for many inspiring comments at all
stages of the review process, in particular for comments to the paper dedicated to period-luminosity relations for Miras in the WISE and Spitzer bands \citep{2020arXiv201212910I}. Remarks raised by the referee led us to explore the subject of multiwavelength variability of Miras (leading, in turn, to writing this paper). PI is partially supported by the {\it Kartezjusz} programme no. POWR.03.02.00-00-I001/16-00 founded by the National Centre for Research and Development, Poland. This work has been supported by the National Science Centre, Poland, via grants OPUS 2018/31/B/ST9/00334 to SK and MAESTRO 2016/22/A/ST9/00009 to IS.  The OGLE project has received funding from the National Science Centre, Poland through MAESTRO grant no. 2014/14/A/ST9/00121. 

\newpage
This publication makes use of data products from the {\it Wide-field Infrared Survey Explorer} (WISE), which is a joint project of the University of California, Los Angeles, and the Jet Propulsion Laboratory/California Institute of Technology, funded by the {\it National Aeronautics and Space Administration} (NASA). This work is based in part on archival data obtained with the Spitzer Space Telescope, which was operated by the Jet Propulsion Laboratory, California Institute of Technology under a contract with NASA.
\end{acknowledgments}

\software{TOPCAT \citep{2005ASPC..347...29T}, R \citep{Rsoftware}, varStar \citep{GPR}, TATRY code \citep{1996ApJ...460L.107S}.}

\balance

\newpage
\bibliography{paper}
\bibliographystyle{aasjournal}

\end{document}